\documentclass[a4paper,fleqn,usenatbib]{mnras}

\usepackage{newtxtext,newtxmath}
\usepackage[T1]{fontenc}
\usepackage{ae,aecompl,multirow,longtable}

\usepackage{graphicx}	
\usepackage{amsmath}	
\usepackage{amssymb}	

\title[Nova search among guest stars]{Cataclysmic variables as possible counterparts of ancient Far Eastern guest stars}

\author[Hoffmann et al.]{
Hoffmann, Susanne M.,$^{1}$\thanks{E-mail: susanne.hoffmann@uni-jena.de (PAF, FSU)}
Vogt, Nikolaus,$^{2}$
\\
$^{1}$Physikalisch-Astronomische Fakult\"at, Friedrich-Schiller-Universit\"at Jena, Germany\\
$^{2}$Instituto de Física y Astronomía, Universidad de Valparaíso, Chile\\
}

\date{Accepted 2020 April 17. Received 2020 April 17; in original form 2020 January 23}

\pubyear{2020}

\begin{document}
\label{firstpage}
\pagerange{\pageref{firstpage}--\pageref{lastpage}}
\maketitle

\begin{abstract}
Continuing our efforts to select possible classical nova candidates among Far Eastern guest stars and to identify them with modern cataclysmic variables (CVs), we present a search for counterparts in 24 promising areas of the sky corresponding to ancient observations between 204 BCE and 1690 CE. These areas have been derived by us in a previous paper. Based on physical entities of the CVs in our areas and reasonable magnitude limits compatible with the distribution of known eruption amplitudes of telescopic classical novae, we present a catalogue of a total of 80 CVs and related targets which could possibly have caused the historical sightings. This list could potentially be reduced by additionally discussing further information more vaguely given in the text. In some cases, we present a detailed discussion of the interpretation of ancient sources confronting them with properties of the brightest CVs in the field.  
In order to estimate whether this list is representative, we discuss the distribution of CV types in our catalogue of counterparts for the historical events. Compared to the entire sky, the surface density of most CV subtypes in our search fields reveals similar values, except for polars and intermediate polars, i.\,e. strongly magnetic CVs, for which a significant excess in our search fields was detected. Finally, we give an outlook towards future research in this topic, and add in an Appendix a complete atlas of the celestial maps of all 24 guest star events, displaying the search areas and locations of CVs within them. 
\end{abstract}




\section{Introduction}
The classical nova phenomenon is one of the most spectacular events in stellar variability and is, thus, a proper candidate to have been observed by historical naked eye astronomers. It refers to a thermonuclear runaway explosion of a hydrogen rich layer on the surface of a white dwarf, accreted previously via the mass transfer from a late-type secondary star. Classical novae are always close semi-detached binaries which are not destroyed by this explosion. The nova phenomenon can repeat many times in the lifetime of this binary configuration, whenever enough hydrogen has been accumulated on the white dwarf to cause instability and trigger a new explosion. Theoretical models as those by \citet{yaron2005} show that the quiescent time between two subsequent eruptions is of the order of $10^4$ to $10^5$ years, and it is still a matter of debate what is happening to the binary during the large time interval between these eruptions.  

Most novae belong to the class of cataclysmic variables (CVs); there are several subtypes of CVs that have similar parameters as size, masses and orbital periods, but not all CVs were observed as novae. Why does their outburst behavior differ so much? Are all CVs novae at certain moments? \citet{vogt1982} suggested that different subtypes of CVs could be transition stages within a large cycle between two subsequent nova eruptions. \citet{shara1984} and \citet{shara1986} modified this picture, introducing a `hibernation' stage in which, as a consequence of the nova eruption, the surface of the secondary CV component would be disconnected for some time from its Roche lobe, implying a break in the mass transfer. Recently, \citet{hillman2020} shed new light on the timescales of these evolutions from feedback-dominated numerical simulations.

A progress in this riddle will only be possible if we could considerably extend the time coverage of CV behavior. There is an attractive possibility to achieve this, if some of the `guest stars' observed by Far Eastern astronomers during the last 2600 years could be identified with modern CVs as remnants of ancient nova eruptions. This approach was successful for eight supernovae, but past attempts to apply it to classical novae revealed only doubtful results, as shown by the recent studies of \citet[hereafter VHT2019]{vogt2019} and \citet{hoffmann2019}. Part of the problems were inadequate interpretations of positions given in old historical sources by modern authors: e.\,g. \citet{hsi,xi+po,pskovskii}, and \citet{steph77} present tables of \textit{point} coordinates per event which is misleading because in most cases only the \textit{areas} are given in the historical record. Working with point coordinates (even after attaching error bars) will lead to wrong results concerning the identification of counterparts. 
\begin{figure*}
    \caption{Naked eye appearance of a bright object in contrast to fainter stars (here: Venus, $-4.6$~mag, next to the central part of the cluster of the Pleiades, stars of 2.9 to 4.3~mag). We do not include any scale in the picture because they shall show the visual impression and there is nothing to measure. Photos are taken with Canon 600D and 300~mm tele objective, without any astronomical preparation, without any filter or any processing in order to display the real impression for the human eye. The pictures show that the bright planet can be described `with rays', `horned', `hairy', or `fuzzy'. The photos were taken in central Europe. The left eight images were taken April 4th to 6th 2020 under normal clear weather conditions. The rightmost picture was taken on April 10th with cirrus clouds showing that atmospheric conditions of the desert or in tropical climate (with sandstorm or humidity) the atmospheric effects become stronger, the bright point source becomes apparently bigger, blurred and has less rays.}
    \label{fig:venus}
	\includegraphics[width=\textwidth]{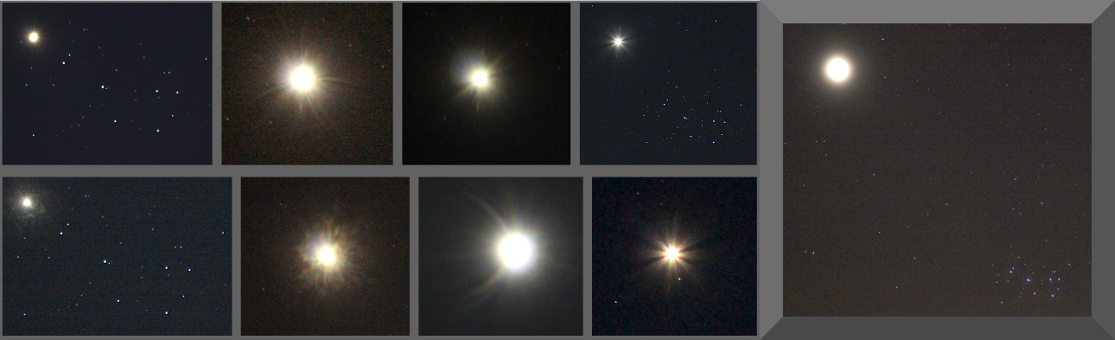}
\end{figure*}

Therefore, for our current project, with clear criteria we selected 185 records (out of several hundred translated in \citet{hsi, ho}, and \citet[129-146]{xu2000}) to derive and test our method presented in this paper. Our selection criteria were: 1) no reported movement or tail (to exclude comets) and 2) not visible only within a certain hour (to exclude meteors). Like Stephenson, Pankenier, Xu and other scholars before us, we neglected wordings like `vapor' or `fuzzy star' or `rays' and such for our selection because a bright object can also appear hairy or with rays for the human eye or `fuzzy' or `washed out' through a dusty or humid atmosphere (cf. Fig.~\ref{fig:venus}). Even the term 'broom star' does not necessarily have to refer to a comet: \citet{hsi} often translates `sweeping star' were `sweeping' or `broom' does not necessarily have the temporal meaning of wiping but can also be interpreted as spacial,\footnote{As many Far Eastern languages the Chinese does not provide \textit{tempora} for verbs. Special thanks to Jesse Chapman, UC Berkeley, for the advise on the time-space-ambivalence of the Chinese.} i.\,e. of a washed-out appearance. In the Almagest, for instance, the star cluster $\omega$~Cen is enrolled in the catalogue as one `nebulous star' and so is the group of three 4~mag-stars in Orion's head. Thus, we neglect wording and consider only reports of long tails with a clear direction as definitely reporting comets. This way, we confirmed the selection of the list of 105 guest star records by \citet[129-146]{xu2000} and the list of 75 likely novae and supernovae by \citet{stephenson} and \citet[p.\,46-49]{steph77} and extended them slightly (cf. Tab.~\ref{tab:list}. However, in none of the cases we can draw a light curve or have any historical follow-up observation to judge what the phenomenon really was. Any final conclusion in this sense would be premature. This requires further work beyond the scope of these other scholars and of our current study. 
\begin{table}  
\label{tab:list}
\caption{Presents of our selected events in earlier nova catalogues: Hsi, Stephenson (and others), and Xu (and Pankenier and others) presented lists of transients they considered as likely novae. In contrast, \citet{ho} gave a catalogue of records not distinguishing between comets and novae/supernovae. Our selected 24 events are all in Ho's collection and show up in the nova lists as displayed with $+$ in this table.}
\begin{tabular} {rccc|rccc} 
 year & Steph.+ & Xu+  & Hsi & year & Steph.+ & Xu+  & Hsi\\
$-203$ & $+$ & $+$  & $+$   &$722$ & $+$ & $+$  & $+$ \\
$-103$ & $-$ & $+$  & $-$ & $840$ & $-$ & $-$  & $+$ \\
 $-47$ & $+$ & $+$  & $+$  & $891$ & $+$ & $+$  & $-$ \\
 $-4$ & $+$ & $-$  & $+$  &$1175$ & $+$ & $+$ & $+$ \\
 $+64$ & $+$ & $+$  & $-$  & $1430$ & $+$ & $+$  & $+$ \\
 $+70$ & $+$ & $+$  & $-$  & $1431$ & $+$ & $+$  & $+$ \\
$+101$ & $+$ & $+$  & $+$  & $1437$ & $+$ & $+$  & $-$ \\
$+329$ & $+$ & $+$  & $-$  &$1461$ & $-$ & $-$  & $+$ \\
$+641$ & $+$ & $+$  & $-$  & $1497$ & $-$ & $+$  & $-$ \\
$+667$ & $-$ & $-$  & $+$  & $1592$ & $+$ & $+$  & $-$ \\
$+668$ & $-$ & $-$  & $+$  & $1661$ &   & $+$  & $-$ \\
$+683$ & $-$ & $-$  & $+$  & $1690$ &   & $+$  & $+$ \\
\end{tabular}
\end{table}

 Aiming to identify modern counterparts, we went further than our predecessors by developing a new method to describe the given celestial positions and use modern databases to search possible candidates of brightening celestial objects. In this (first) paper, we present the search for novae. In order to define reliable areas in the sky for a search of modern counterparts \citep[hereafter HVP2020]{hoffmannVogtProtte}, we reconsidered positions and sizes of ancient asterisms and constellations mentioned in old records of guest stars. Based on the original descriptions of the above mentioned 185 old events from the last $\sim2600$ years, HVP2020 selected some cases to test this further analysis strategy. For this shortlist of `higher priority guest stars', we applied three selection criteria as stated in HVP2020:
 \begin{enumerate} 
  \item Transients with a given duration (but, then, the position could be imprecisely known, i.\,e. only an area given, somewhere in a constellation), because these cases are most promising to derive conclusions on binary evolution. 
  \item \textbf{or} Transients with a well defined position given, i.\,e. next to a single star (but possibly no duration given) because these cases have small search fields and the identification of a CV counterpart appears more likely. 
  \item \textbf{and} Not already suggested as supernova or comet.  
 \end{enumerate} 
 The logical rule is [(\textit{i}) $\vee$ (\textit{ii})] $\wedge$ (\textit{iii}). The result was a shortlist of 24 events (cf. Tab.~\ref{tab:list}) which could refer to classical nova eruptions and could -- in case of a found counterpart -- possibly contribute to modern questions of research, e.\,g. \citep{tappert2012,pat2013,miszalski2016}. The present study describes an adequate strategy to search for modern counterparts of these historical phenomena and gives a list of possible identifications among catalogued CVs (Section~2). The most promising candidates for classical novae are discussed in more detail in Section~3. Finally, we add some statistical considerations in Section~4 and discuss our results in Section~5, including some perspectives and requirements for further studies.    

\section{Search methods and strategies}\label{sec:CVlist}
A postnova system today has the apparent brightness $m$ computed as sum of the outburst peak brightness $m_\textrm{peak}$ and the amplitude $\Delta m$ of the nova eruption: $m_\textrm{postnova} =  m_\textrm{peak} + \Delta m$. As the amplitude of a certain historical event is, of course, unknown, we consider the typical distribution of amplitudes of novae in the past one and a half century: Fig.~1 in VHT2019 suggests typical amplitudes of 11 to 13~mag and the rare maximum amplitude of 16~mag (example: V1500 Cyg). This means, that it is possible but unlikely that a nova eruption has an amplitude of 16~mag and that $\sim90$\,\% of all novae show amplitudes of 13~mag or smaller. In order to generate a shortlist of likely candidates, we decided to assume a typical nova amplitude of $\Delta m =13$~mag. 

The value for the peak brightness is much more difficult to estimate: Unfortunately, for historical novae we typically do not know their peak brightness. Only in rare cases, it is preserved whether the new star (or guest star) was `small' or `large', faint or bright. Yet, even in those cases, we cannot know whether the writer estimated a 3 or 4~mag star as bright or faint. In texts of Greek mathematical astronomy, for instance, this terminology depends on the background brightness in this area and the vicinity of bright stars \citep[p.\,91-92 and p.\,191-194]{smh2017}. For Far Eastern texts we do not have enough data to prove this but as this is a common principle of human physiology and cognition it suggests a similar effect \citep[Fig.\,3]{hoffmannVogtProtte}. 

For the observed brightness of the historical nova we only have the information that it was visible to the naked eye. Hence, we could check historical star catalogues to which limit they contain stars: Fig.~2 in VHT2019 displays the distribution of stars of the historical star catalogues. Until the 4th magnitude both catalogues are (almost) complete \citep[map on p. 247]{smh2017} but the Chinese catalogue reaches further south and contains more stars. It lists many more stars of 5th magnitude than the Greek one but it is still far from being complete. The distribution of the stars of the Chinese catalogue of 5th and fainter magnitudes on the celestial map shows clearly that they were registered preferably outside of the Milky Way (cf. HVP2020, Fig.~3) while the distribution of 4~mag stars seems to be smoother all over the map. This suggests that a new star until 4~mag would have been realized by the ancient naked eye astronomers with considerable likelihood\footnote{This hypothesis derived from historical catalogues is confirmed by the observational experience by one of us (SH).} while a new star of 5~mag would likely have been overlooked in the Milky Way and registered outside the Milky Way only in peculiar cases. For instance, it is more likely to realize a 5~mag star if it appears next to a 3~mag star (e.\,g. in case of an outburst of BK Lyn up to 5~mag because the star is situated next to $\alpha$ Lyn with $V = 3.14$~mag). 

Applying these thoughts to our goal to estimate the postnova brightness by $m_\textrm{postnova} =  m_\textrm{peak} + \Delta m$ we find that postnovae with peak brightness $< 4$~mag and eruption amplitude $<13$~mag amplitude reveals the limit $<17$~mag for a relatively high detection probability by ancient astronomers (the brighter, the more likely). 5~mag $+ 16$~mag $= 21$~mag seems to be the faintest limit for the brightness a system might have today to be considered as a candidate for a naked eye nova. However, the magnitudes between 21~mag and 17~mag are much less likely for postnova counterparts of naked eye novae than the brighter CVs.   

In practice, some additional considerations have to be applied: Considering the behavior of cataclysmic binary systems in more detail, we find that `a current brightness' does not really exist: All CVs show short-term variability such as flickering and variations with the orbital period as eclipses,  humps, irradiation or ellipsoidal deformation of the secondary component. These short-term variations are typically of the order of $\pm0.5\dots1$~mag except eclipses, which could be rather deep, up to $2-3$~mag. Therefore, we limit our search to $V = 18$~mag, adding one magnitude to the above mentioned limit $V = 17$~mag taking into account the short-term variability. The final limit of $V = 18$~mag will be applied on dwarf novae at minimum light (quiescent state); but for nova-like stars, magnetic CVs (polars and intermediate polars) and other CVs of unknown subclass we applied this limit at maximum light because in most cases it is impossible to calculate a reliable value of its mean brightness, and especially the rather frequent subclass of VY Scl stars is characterized by temporary drops in their brightness. Modern classical novae are also possible identifications, because they could be recurrent novae, with an ancient and a later `modern' outburst event. However, this will only be possible if their eruption maximum reached naked eye brightness. On the other hand, even for modern rapidly fading novae the real maximum could have been missed; therefore, we consider only novae with peak brightness $m < 7$~mag as possible counterparts. Finally, low-mass X-ray binaries could also be possible candidates, since among them there are modern X-ray novae, for instance GU Mus $=$ Nova Muscae 1991, a close binary with a black hole as primary component \citep{wu2016}.  

 As a caveat, we want to stress that we consider the faintest postnova magnitude of a historical sighting nowadays of the order $\sim21$~mag. In this study, we present the more likely brighter CVs down to a limit of $18$~mag. This criterion will possibly not find the solution for some historical novae; e.\,g. WY Sge (Nova 1783) which has a current V magnitude of $\sim18.7$~mag \citep{shara1984}. This example again supports our selection criterion for CVs of high likelihood and leaves room for further investigations of a bigger sample of observational targets.  

\begin{table*}
	\centering
	\begin{minipage}{\textwidth}
	\caption{Result of the search of CVs and low-mass X-ray binaries (in 22 areas according to Table~3 of HVP2020) as possible identifications of stellar remnants from Far Eastern guest star nova eruptions. These CVs match our first criterion being brighter than 18~mag but have to be considered case-by-case considering the historical record. Columns $4-7$ give equatorial coordinates in degrees (epoch 2000) and the modern constellation of each target. Columns $8-10$ list subtype, orbital period and magnitude range, resp. according to the VSX AAVSO catalogue. In column~11 target stars $V\leq16$~mag are marked. The comment in column~12 refers to the evaluating of additional information given in the text as shown in the text: see Tab.~\ref{tab:novaCandDiscuss}.}
	\label{tab:tab1}
	\scriptsize
	\begin{tabular}{cccp{2.88cm}cc cccccp{2.55cm}} 
		\hline
		index	&id	&year	&Name	&RA/ $\degr$	&DE/ $\degr$	&Const.	&Type	&Period	&Mag	&$<16$?	&comment\\
		 &&&&&&&&&&& (see case discussion) \\
		\hline
 1. & 1. & -203. & \text{RX J1404.4+1723} 			& 211.122 & 17.3997 & \text{Boo} & \text{UG} & \text{--} & \text{14.0 - 17.5 CV} & \text{} & \text{} \\
 2. & 2. & -203. & \text{AB Boo} 					& 211.696 & 20.8069 & \text{Boo} & \text{N:} & \text{--} & \text{4.5 - ? V} & + & \text{} \\
 3. & 3. & -203. & \text{SDSS J143209.78+191403.5} 	& 218.041 & 19.2342 & \text{Boo} & \text{NL/VY+E} & 0.1174 & \text{18.0 - $<$20.7 CV} & \text{} & \text{} \\
 4. & 1. & -47. & \text{OGLE-BLG-DN-1055} & 278.191 & -24.2712 & \text{Sgr} & \text{UG} & \text{--} & \text{15.4 - 19.0 Ic} & \text{} & \text{too faint} \\
 5. & 2. & -47. & \text{OGLE-BLG-DN-1057} & 279.103 & -23.9099 & \text{Sgr} & \text{UG} & \text{--} & \text{15.2 - 17.3 Ic} & \text{} & \text{too faint} \\
 6. & 1. & 64. & \text{V0379 Vir} 					& 183.039 & 1.6075 & \text{Vir} & \text{AM} & 0.061408 & \text{17.9 (0.10) R} & \text{} & \text{too faint} \\
 7. & 1. & 70. & \text{SDSS J083751.00+383012.5} 	& 129.462 & 38.5033 & \text{Lyn} & \text{AM} & 0.1325 & \text{16.9 - 18.9 CV} & \text{} & \text{} \\
 8. & 2. & 70. & \text{BP Lyn} 						& 135.787 & 41.2964 & \text{Lyn} & \text{E+NL} & 0.152813 & \text{14.19 - 14.33 B} & + & \text{} \\
 9. & 3. & 70. & \text{BK Lyn} 						& 140.047 & 33.945 & \text{Lyn} & \text{UGER} & 0.07498 & \text{14.3 - 16.5 V} & + & \text{} \\
 10. & 4. & 70. & \text{SDSS J092122.84+203857.1} 	& 140.345 & 20.6489 & \text{Cnc} & \text{AM} & 0.0584999 & \text{16.8 - 19.2 r'} & \text{} & \text{} \\
 11. & 5. & 70. & \text{Nova Leo 1612} 				& 144.329 & 15.22 & \text{Leo} & \text{N:} & \text{--} & \text{4 - ? v} & + & \text{} \\
 12. & 6. & 70. & \text{SDSS J094002.56+274942.0} 	& 145.011 & 27.8283 & \text{Leo} & \text{UG} & 0.16352 & \text{14.6 - 18 CR} & \text{} & \text{} \\
 13. & 7. & 70. & \text{HY Leo}						& 146.644 & 13.8492 & \text{Leo} & \text{DQ} & 0.171875 & \text{15.8 - 20.0 CV} & + & \text{} \\
 14. & 8. & 70. & \text{X Leo} 						& 147.756 & 11.8753 & \text{Leo} & \text{UGSS} & 0.1644 & \text{11.8 - 17.2 V} & \text{} & \text{} \\
 15. & 9. & 70. & \text{RX J0953.1+1458} 			& 148.284 & 14.9767 & \text{Leo} & \text{AM} & 0.072049 & \text{17.2 - 19.1 CV} & \text{} & \text{} \\
 16. & 10. & 70. & \text{SDSS J100658.40+233724.4} 	& 151.743 & 23.6233 & \text{Leo} & \text{UG+E} & 0.185913 & \text{14.7 - 20.5 V CV} & \text{} & \text{} \\
 17. & 11. & 70. & \text{SDSS J101421.55+063857.7}	& 153.59 & 6.64917 & \text{Leo} & \text{CV} & \text{--} & \text{14.00 - 14.08 CV} & + & \text{} \\
 18. & 12. & 70. & \text{GG Leo} 					& 153.895 & 9.07806 & \text{Leo} & \text{AM} & 0.0554719 & \text{15.26 V - 19.2 CV} & \text{} & \text{} \\
 19. & 13. & 70. & \text{MV Leo} 					& 157. & 21.8036 & \text{Leo} & \text{UGZ+E} & 0.14608 & \text{15.2 - 17.7 CV} & \text{} & \text{} \\
 20. & 1. & 101. & \text{BK Lyn} 					& 140.047 & 33.945 & \text{Lyn} & \text{UGER} & 0.07498 & \text{14.3 - 16.5 V} & + & \text{} \\
 21. & 2. & 101. & \text{YZ LMi}					& 141.661 & 36.4006 & \text{LMi} & \text{UGSU+E/IBWD} & 0.01966 & \text{16.5 - 20.5 CV} & \text{} & \text{} \\
 22. & 3. & 101. & \text{SDSS J100658.40+233724.4} 	& 151.743 & 23.6233 & \text{Leo} & \text{UG+E} & 0.185913 & \text{14.7 - 20.5 V CV} & \text{} & \text{} \\
 23. & 1. & 329. & \text{DW UMa} 					& 158.47 & 58.7817	 & \text{UMa} & \text{NL/VY+E} & 0.136607 & \text{13.6 - 18.0 V} & \text{} & \text{2-3${}^{\circ}$ away } \\
 24. & 2. & 329. & \text{IY UMa} 					& 160.986 & 58.1253 & \text{UMa} & \text{UGSU+E} & 0.073909 & \text{13.0 - 18.9 V} & \text{} & \text{} \\
 25. & 3. & 329. & \text{SDSS J113215.50+624900.4} 	& 173.065 & 62.8167 & \text{UMa} & \text{CV} & \text{--} & \text{15.1 - 19.5 CV} & \text{} & \text{No; in asterism of jugdes } \\
 26. & 4. & 329. & \text{GP CVn} 					& 186.92 & 51.6569 & \text{CVn} & \text{UGSU+E} & 0.0629504 & \text{14.3 - 19.1 g' CV} & \text{} & \text{4-5${}^{\circ}$ away} \\
 27. & 5. & 329. & \text{EV UMa} 					& 196.974 & 53.8583 & \text{UMa} & \text{AM} & 0.0553389 & \text{17 - 21 V} & \text{} & \text{} \\
 28. & 6. & 329. & \text{SBS 1316+577A} 			& 199.503 & 57.4678 & \text{UMa} & \text{CV:} & \text{--} & \text{16.5 - ? B} & \text{} & \text{2-3${}^{\circ}$ away } \\
 29. & 7. & 329. & \text{V0496 UMa} 				& 200.267 & 56.1658 & \text{UMa} & \text{AM} & 0.063235 & \text{16.1 - 20.0 V} & \text{} & \text{} \\
 30. & 8. & 329. & \text{UX UMa} 					& 204.171 & 51.9136 & \text{UMa} & \text{EA/WD+NL} & 0.196671 & \text{12.57 - 14.15 V} & + & \text{} \\
 31. & 9. & 329. & \text{V0355 UMa} 				& 204.921 & 48.7908 & \text{UMa} & \text{UGWZ+ZZ/GWLIB} & 0.05729 & \text{10.3 - 17.7 V} & \text{} & \text{} \\
 32. & 10. & 329. & \text{CT Boo} 					& 212.087 & 53.5111 & \text{Boo} & \text{NL:} & \text{--} & \text{17 - 18.7: B} & \text{} & \text{4-5${}^{\circ}$ away} \\
 33. & 1. & 641. & \text{SDSS J122405.58+184102.7}	 & 186.023 & 18.6839 & \text{Com} & \text{NL/VY} & \text{--} & \text{15.5 - 20.0 CV} & \text{} & \text{} \\
 34. & 2. & 641. & \text{SDSS J123255.11+222209.4} 	& 188.23 & 22.3692 & \text{Com} & \text{CV} & \text{--} & \text{17.6 - 18.1 CR} & \text{} & \text{too faint} \\
 35. & 3. & 641. & \text{IR Com} 					& 189.884 & 21.135 & \text{Com} & \text{UG+E} & 0.0870386 & \text{13.4 - 18.5 p} & \text{} & \text{} \\
 36. & 1. & 667. & \text{2MASS J04030790+3110038} 	& 60.7829 & 31.1675 & \text{Per} & \text{NL:} & \text{--} & \text{15.5 - 16.8 CV} & + & \text{} \\
 37. & 2. & 667. & \text{2MASS J04132921+3116279} 	& 63.3718 & 31.2744 & \text{Per} & \text{CV} & \text{--} & \text{16.3 - 18.1 CV} & \text{} & \text{too faint} \\
 38. & 3. & 667. & \text{ASASSN-16pm} 				& 64.353 & 22.2561 & \text{Tau} & \text{AM:+E} & 0.153988 & \text{14.1 - 21.0 CV} & + & \text{} \\
 39. & 4. & 667. & \text{V0518 Per} 				& 65.4283 & 32.9075 & \text{Per} & \text{LMXB/BHXB/XN} & 0.21216 & \text{13.15 - $<$22.4 V} & \text{} & \text{} \\
 40. & 1. & 668. & \text{NSV 1586} 					& 66.0749 & 47.87 & \text{Per} & \text{UG} & \text{--} & \text{15.1 - $<$16.7 V} & + & \text{} \\
 41. & 2. & 668. & \text{FY Per} 					& 70.486 & 50.71 & \text{Per} & \text{NL/VY} & 0.2584 & \text{11.9 - 14.5 V} & + & \text{} \\
 42. & 3. & 668. & \text{V0392 Per} 				& 70.839 & 47.3569 & \text{Per} & \text{NA+UG} & \text{--} & \text{6.3 - 16.9 V} & \text{} & \text{no (not visible in twilight)} \\
 43. & 4. & 668. & \text{ASASSN-15rs} 				& 71.6403 & 48.9653 & \text{Per} & \text{UGSU} & \text{--} & \text{14.5 - 17.5 V} & \text{} & \text{} \\
 44. & 5. & 668. & \text{IGR J04571+4527} 			& 74.2791 & 45.4633 & \text{Aur} & \text{DQ} & \text{0.258:} & \text{16.7 - 17.4 CR} & \text{} & \text{too faint} \\
 45. & 6. & 668. & \text{MGAB-V627} 				& 78.4684 & 50.5353 & \text{Aur} & \text{NL/VY} & \text{--} & \text{18.1 - 20.4 r} & \text{} & \text{too faint} \\
 46. & 7. & 668. & \text{ASASSN-17oz} 				& 91.979 & 45.205 & \text{Aur} & \text{UG} & \text{--} & \text{13.7 - 16.6 V} & + & \text{} \\
 47. & 8. & 668. & \text{SS Aur} 					& 93.3435 & 47.7403 & \text{Aur} & \text{UGSS} & 0.1828 & \text{10.3 - 16.8 V} & + & \text{} \\
 48. & 9. & 668. & \text{V0552 Aur} 				& 93.5409 & 45.5022 & \text{Aur} & \text{DQ:} & 0.060868 & \text{11.2 - 14.5 p} & + & \text{} \\
 49. & 10. & 668. & \text{2MASS J06324931+4529419} 	& 98.2055 & 45.4947 & \text{Aur} & \text{CV} & \text{--} & \text{16.2 - 18.0 CV} & \text{} & \text{too faint} \\
 50. & 1. & 683. & \text{MGAB-V627} 				& 78.4684 & 50.5353 & \text{Aur} & \text{NL/VY} & \text{--} & \text{18.1 - 20.4 r} & \text{} & \text{too faint} \\
 51. & 2. & 683. & \text{DDE 38} 					& 82.905 & 48.4086 & \text{Aur} & \text{NL:} & \text{--} & \text{16.7 r$\&\#$039; - 18.4 R} & \text{} & \text{too faint} \\
 52. & 3. & 683. & \text{ASASSN-14gy}				 & 84.8116 & 45.4739 & \text{Aur} & \text{UG} & \text{--} & \text{15.74 - $<$17.4 V} & \text{} & ? \\
 53. & 4. & 683. & \text{MGAB-V330} 				& 87.1278 & 53.7339 & \text{Aur} & \text{AM+E} & \text{--} & \text{16.3 - 20.5 g} & \text{} & \text{too faint} \\
 54. & 1. & 722. & \text{GALEX J003535.7+462353} 	& 8.89883 & 46.3978 & \text{And} & \text{UG+E} & 0.172275 & \text{14.5 - 18.5 V} & \text{} & \text{} \\
 55. & 2. & 722. & \text{HV And} 					& 10.2308 & 43.4164 & \text{And} & \text{NL/VY} & 0.1403 & \text{15.2 - $<$18.7 G g} & \text{} & \text{} \\
 56. & 3. & 722. & \text{ASASSN-13bz} 				& 11.1153 & 43.7117 & \text{And} & \text{UG} & \text{--} & \text{16.1 - $<$17 V} & \text{} & \text{too faint} \\
 57. & 4. & 722. & \text{V0515 And} 				& 13.8328 & 46.2158 & \text{And} & \text{DQ} & \text{0.10935:} & \text{14.7 - 16.0 V} & + & \text{} \\
 58. & 5. & 722. & \text{DDE 169} 					& 14.768 & 46.2467 & \text{And} & \text{UG} & \text{--} & \text{14.5 - 17.9 V} & \text{} & \text{} \\
 59. & 6. & 722. & \text{V0723 Cas} 				& 16.2723 & 54.0111 & \text{Cas} & \text{NB} & 0.693265 & \text{7.08 - $<$18.0 V} & \text{} & \text{too faint} \\
 60. & 7. & 722. & \text{2MASS J01074282+4845188} 	& 16.9285 & 48.755 & \text{Cas} & \text{EA+NL} & 0.193598 & \text{15.0 - 17.2 V} & \text{} & \text{} \\
 61. & 8. & 722. & \text{HT Cas} 					& 17.554 & 60.0767 & \text{Cas} & \text{UGSU+E} & 0.0736472 & \text{12.6 - 19.32 V} & \text{} & \text{} \\
 62. & 9. & 722. & \text{NSV 15272} 				& 19.6136 & 63.6958 & \text{Cas} & \text{CV:} & \text{--} & \text{10.0 - 17.7 B} & \text{} & \text{} \\
 63. & 10. & 722. & \text{IPHAS J013031.90+622132.4} & 22.6329 & 62.3589 & \text{Cas} & \text{CV} & 0.130062 & \text{16.9 - ? r'} & \text{} & \text{too faint} \\
 64. & 11. & 722. & \text{KU Cas} 					& 22.76 & 57.9033 & \text{Cas} & \text{UGSS} & \text{(60)} & \text{13.3 - 18.0 p} & \text{} & \text{} \\
 65. & 12. & 722. & \text{MGAB-V576} 				& 26.3728 & 58.5681 & \text{Cas} & \text{UG+E} & 0.0911617 & \text{16.4 - 21.0 g} & \text{} & \text{too faint} \\
 66. & 1. & 840. & \text{CSS 110921:000721+200722} 	& 1.83663 & 20.1225 & \text{Peg} & \text{UG} & \text{--} & \text{13.6 - 17.6 CV} & \text{} & \text{} \\
 67. & 2. & 840. & \text{V0405 Peg} 				& 347.455 & 21.5881 & \text{Peg} & \text{CV} & 0.177646 & \text{15.1 - 17.1 V} & \text{} & \text{} \\
 68. & 3. & 840. & \text{Swift J2319.4+2619} 		& 349.877 & 26.2553 & \text{Peg} & \text{AM} & \text{--} & \text{15.6 - 20.6 CR} & \text{} & \text{} \\
 69. & 4. & 840. & \text{IP Peg} 					& 350.786 & 18.4164 & \text{Peg} & \text{UG+E} & 0.158206 & \text{12 - 18.6 B} & \text{} & \text{} \\
 70. & 5. & 840. & \text{V0598 Peg} 				& 353.358 & 15.3728 & \text{Peg} & \text{DQ} & 0.05771 & \text{17.8 - 19.3 CV} & \text{} & \text{too faint} \\
 71. & 6. & 840. & \text{V0378 Peg} 				& 355.018 & 30.2964 & \text{Peg} & \text{NL} & 0.13858 & \text{13.6 - 14.3 CV} & \text{} & \text{} \\
 72. & 1. & 891. & \text{ASASSN-15dl} 				& 245.502 & -22.6880 & \text{Sco} & \text{UG} & 0.2288 & \text{14.3 - 17.9 CV} & \text{} & \text{} \\
 73. & 2. & 891. & \text{U Sco} 					& 245.628 & -17.8785 & \text{Sco} & \text{NR+E} & 1.23055 & \text{7.5 - 19.3 V} & \text{} & \text{} \\
 74. & 1. & 1175. & \text{SDSS J150137.22+550123.4} & 225.405 & 55.0231 & \text{Dra} & \text{UGSU+E} & 0.0568413 & \text{16.4 - 19.4 CR} & \text{} & \text{too faint} \\
 75. & 2. & 1175. & \text{SDSS J153817.35+512338.0} & 234.572 & 51.3939 & \text{Boo} & \text{AM} & 0.06466 & \text{17.5 - 19.1 CV} & \text{} & \text{too faint} \\
	\end{tabular}
	\end{minipage} 
\end{table*} %
\begin{table*}
	\centering
	\begin{minipage}{\textwidth}
	\contcaption{ }
	\scriptsize
	\begin{tabular}{cccp{3cm}cccccccp{3cm}} 
		\hline
		index	&id	&year	&Name	&RA/ $\degr$	&DE/ $\degr$	&Const.	&Type	&Period	&Mag	&$<16$?	&comment\\
		\hline
 76. & 1. & 1430. & \text{BG CMi} 							& 112.871 & 9.93972 & \text{CMi} & \text{DQ} & 0.13475 & \text{13.9 - 15.6 V} & + & \text{more likely} \\
 77. & 2. & 1430. & \text{MASTER OT J074940.48+071557.1} 	& 117.419 & 7.26583 & \text{CMi} & \text{UG} & \text{--} & \text{15.3 - $<$17.9 CR} & \text{} & \text{} \\
 78. & 1. & 1431. & \text{ASASSN-14er} 						& 66.0052 & 2.69694 & \text{Tau} & \text{UG} & \text{--} & \text{15.34 - $<$17.1 V} & \text{} & \text{} \\
 79. & 2. & 1431. & \text{IW Eri} 							& 66.48 & -19.7585 & \text{Eri} & \text{AM} & 0.0605 & \text{16.1 - 20.2 V} & \text{} & \text{} \\
 80. & 3. & 1431. & \text{CSS 110113:043112-031452} 		& 67.8019 & -3.2477 & \text{Eri} & \text{UGSU+E} & 0.06603 & \text{14.9 - 20 CV} & \text{} & \text{} \\
 81. & 4. & 1431. & \text{CSS 100218:043829+004016} 		& 69.6213 & 0.671111 & \text{Tau} & \text{UG+E} & 0.06546 & \text{16.4 - 19 CV} & \text{} & \text{} \\
 82. & 5. & 1431. & \text{BF Eri} 							& 69.8748 & -4.5997 & \text{Eri} & \text{UG} & 0.27088 & \text{12.3 - 15.2 V} & + & \text{most likely 1} \\
 83. & 6. & 1431. & \text{ASASSN-16dl} 						& 70.3556 & -16.9425 & \text{Eri} & \text{UG} & \text{--} & \text{14.13 - $<$16.3 V} & \text{} & \text{} \\
 84. & 7. & 1431. & \text{KT Eri} 							& 71.9758 & -10.1786 & \text{Eri} & \text{NA+ZAND:+E:} & 737. & \text{5.4 - 16.5: V} & \text{} & \text{too faint} \\
 85. & 8. & 1431. & \text{CSS 110114:044903-184129} 		& 72.2612 & -18.6914 & \text{Eri} & \text{UGSS+E} & 0.15554 & \text{14.9 - 18.3 G CV} & \text{} & \text{} \\
 86. & 9. & 1431. & \text{HY Eri} 							& 75.4432 & -3.989 & \text{Eri} & \text{AM+E} & 0.118969 & \text{16.0 - 22.7 V} & \text{} & \text{} \\
 87. & 10. & 1431. & \text{AQ Eri} 							& 76.5546 & -4.1353 & \text{Eri} & \text{UGSU} & 0.06094 & \text{12.5 - 17.8 V} & \text{} & \text{most likely 2} \\
 88. & 1. & 1437. & \text{V1101 Sco} 						& 256.435 & -36.42297 & \text{Sco} & \text{LMXB} & 0.9375 & \text{18.3 - 19.3 V} & \text{} & \text{} \\
 89. & 1. & 1461. & \text{V2951 Oph} 						& 262.527 & 3.63861 & \text{Oph} & \text{AM} & 0.0834785 & \text{16.8 - 20.8 G R} & \text{} & \text{} \\
 90. & 2. & 1461. & \text{4PBC J1740.7+0603} 				& 265.191 & 6.06417 & \text{Oph} & \text{CV} & \text{--} & \text{15.9 - ? V} & \text{} & \text{} \\
 91. & 3. & 1461. & \text{V0380 Oph} 						& 267.557 & 6.09139 & \text{Oph} & \text{NL/VY} & 0.153477 & \text{14.3 V - 19.0 CV} & \text{} & \text{} \\
 92. & 1. & 1592. & \text{MGAB-V496}						 & 1.47825 & 56.5594 & \text{Cas} & \text{NL/VY} & \text{--} & \text{17.9 - $<$21.1 g} & \text{} & \text{} \\
 93. & 2. & 1592. & \text{MGAB-V239} 						& 4.91392 & 54.0919 & \text{Cas} & \text{NL+E} & \text{--} & \text{18.0 - 20.0: r} & \text{} & \text{} \\
 94. & 3. & 1592. & \text{V0592 Cas} 						& 5.21767 & 55.7044 & \text{Cas} & \text{NL} & 0.115063 & \text{12.79: (0.4:) V} & + & \text{} \\
 95. & 4. & 1592. & \text{V1033 Cas} 						& 5.74017 & 61.6853 & \text{Cas} & \text{DQ} & 0.16804 & \text{16.1 - 16.9 V} & \text{} & \text{} \\
 96. & 5. & 1592. & \text{V0410 Cas} 						& 5.86525 & 61.7747 & \text{Cas} & \text{NL} & \text{--} & \text{15.5 - 18.0 p} & \text{} & \text{} \\
 97. & 6. & 1592. & \text{V0709 Cas} 						& 7.20333 & 59.2894 & \text{Cas} & \text{DQ} & 0.222204 & \text{14.75 - 15.35 B} & + & \text{} \\
 98. & 7. & 1592. & \text{V1037 Cas} 						& 7.26271 & 59.5717 & \text{Cas} & \text{LMXB/XP} & 0.1025 & \text{17.4 - 23.3 R} & \text{} & \text{} \\
 99. & 8. & 1592. & \text{ASASSN-18aan} 					& 11.5335 & 62.1681 & \text{Cas} & \text{UGSU+E} & 0.14944 & \text{15.2 - 17.1 V} & \text{} & \text{} \\
 100. & 1. & 1661. & \text{PW Aqr} 							& 312.574 & -5.6075 & \text{Aqr} & \text{AM+E} & 0.0654246 & \text{17.75 - 19.28 V} & \text{} & \text{too faint} \\
 101. & 2. & 1661. & \text{Gaia17avi} 						& 314.272 & -7.4072 & \text{Aqr} & \text{CV$|$QSO} & \text{--} & \text{17.8 - 20.8 CV} & \text{} & \text{too faint} \\
 102. & 1. & 1690. & \text{V1595 Sgr} 						& 275.143 & -33.3008 & \text{Sgr} & \text{UG} & \text{--} & \text{14.9 - 16.8: V} & + & \text{} \\
 103. & 2. & 1690. & \text{V0909 Sgr} 						& 276.468 & -35.0240 & \text{Sgr} & \text{NA+E} & 0.14286 & \text{6.8 - 20.4 V p} & \text{} & \text{too faint} \\
 104. & 3. & 1690. & \text{V2572 Sgr} 						& 277.903 & -32.5995 & \text{Sgr} & \text{NA} & \text{--} & \text{6.5 - 17.8 R B} & \text{} & \text{too faint} \\
 105. & 4. & 1690. & \text{ASASSN-17nd} 					& 278.897 & -34.1967 & \text{Sgr} & \text{UG} & \text{--} & \text{15.0 - 17.3 V} & \text{} & \text{} \\
 106. & 5. & 1690. & \text{ASASSN-19rb} 					& 279.164 & -34.5179 & \text{Sgr} & \text{UG} & \text{--} & \text{15.5 - 18.0 V g} & \text{} & \text{} \\
 107. & 6. & 1690. & \text{V4745 Sgr} 						& 280.011 & -33.4488 & \text{Sgr} & \text{NB/DQ:} & 0.20782 & \text{7.3 - $<$17.0 V} & \text{} & \text{too faint} \\
		\hline
	\end{tabular}
	\end{minipage}
\end{table*}
Our search was performed for 24 ancient guest star events that could have been classical nova eruptions, according to the above-mentioned criteria. The corresponding areas are listed in HVP2020 (there: Table~3) as circles around Hipparcos stars. We used the grouping `CVs' in the Variable Star Index of the American Association of Variable Star Observers (VSX, AAVSO: \citet{watson}) in November 2019 for this search, and selected among the targets in these areas different CV subtypes: dwarf novae (subtype UG), magnetic CVs (subtypes AM and DQ), and nova-like stars (subtypes NL). On the other hand, there are also CVs in the VSX with unknown subtype (marked as `CV'), and, outside the grouping of `CVs', the low-mass X-ray binaries (class LMXB). In two of the 24 events (namely $+1497$ in UMi and $-103$ around $\gamma$~Boo) no candidates were found. Therefore, Table~\ref{tab:tab1} contains the results for the remaining 22 events with a total of 107 valid candidates. In average, we found about 5 candidates per event, ranging from 1 (for events 64 and 1437) to 13 (for event 70). 

\section{Statistical considerations}\label{sec:stat}
In Table~\ref{tab:Tab4} we display the observed durations, the number of search circles, and the size of the areas they cover. The results of this search for CVs are listed in Tab.~\ref{tab:tab1}. In Tab.~\ref{tab:Tab5} we present the numbers of CVs expected from the total counts per subtype in the VSX catalogue (applying the same limits in magnitude for each subtype, as explained in the previous section) and compare them to the subtype distribution in Tab.~\ref{tab:tab1}. The total number fits perfectly the expectations, and there are only insignificant differences in most of the individual subtypes. However, for the two magnetic CV subtypes AM and DQ there is, for each of them, a remarkable excess in our sample, significantly near $3~\sigma$, as compared to the numbers expected from the VSX. Combining AM and DQ we get 23 cases, revealing an excess of ($98 \pm 21$\,\%), i.\,e. a signal significant at $4.7~\sigma$. 

Although the large majority of targets in our shortlist of 107 stars in Tab.~\ref{tab:tab1} are dwarf novae, this is not the case for the group of the brightest ones (marked as $<16$) which are more likely candidates for naked-eye visibility: Only one of them (BF~Eri) is a dwarf nova, all remaining stars are nova-like, magnetic or unclassified CVs. This could be a first hint that most dwarf novae might not be valid candidates for a correct identification as nova remnants among Far Eastern guest stars. Based on known nova amplitudes, we do expect in average relatively bright counterparts for ancient novae visible at naked-eye brightness. 

\begin{table}
	\centering
	\caption{Solid angle sizes in square degrees of the search areas of the 25 Far Eastern guest star events in Table 1, according to Table 3 of HVP2020. The duration (in days) refers to the total visibility time in the guest star, N$_\circ$ refers to the number of search circles for each event, N$_\star$ to the number of remaining candidates. For further explanations see text above this section.}
	\label{tab:Tab4}
	\begin{tabular}{rrrrr} 
		\hline
Event      &Duration     &N$_\circ$        &  Area  	 \\
           &         (d) &         & (square degrees)   \\
$-203$            & $>10$       & 1 &  78.5  	 \\
$-103 $            &  $-$        & 1 &  28.3  	 \\   
$-47 $             &  $-$      & 1 &   50.3 	 \\
$-4 $              &   $-$     & 3 &   41.0 	 \\
64               &  75       & 1 &    78.5 		 \\
70               &  48       & 5 &   511.8 		 \\
101              &  $-$        & 2 &     56.5 	 \\
329              & 23        & 5 &  263.1 		 \\
641              & 25        & 2 &     57.3 	 \\
667              & $-$       & 1 &    153.9 	 \\
668              & $-$       & 2 &    106.8 	 \\
683              & $-$       & 1 &      78.5 	 \\
722              &   5       & 5 &    162.3 	 \\
840              & $-$       & 1 &    254.5 	 \\
891              & $-$       & 1 &     28.3 	 \\
1175             &  5        & 1 &  201.1 		\\
1430             & 26        & 2 & 35.34 		 \\
1431             & 15        & 5 &  274.1 		 \\
1437             & 14        & 2 & 78.5 		 \\
1461             &   3       & 1 &   50.3		 \\
1497             & $-$       & 1 &   28.3 		 \\
1592-1          &450         & 2 &  3.1 		 \\ 
1592-2         & 120         & 2 & 78.5  		 \\ 
1661            &   19       & 1 &   28.3 		 \\
1690            &   2        & 1 &   28.3 		 \\
               & In total   &   &2698.9    \\      
		\hline
	\end{tabular}
\end{table}

There is another remarkable fact: The VSX catalogue of AAVSO lists many doubtful novae as `N:', but only 3 of them have reported naked eye brightness at maximum light: AB Boo 1877 (4.5~mag), N Leo 1612 (4.0~mag) and NSV 3846 Pup 1673 (3.0~mag). The next fainter doubtful nova has a peak brightness of 7.5~mag. Two of these 3 possible naked eye sightings of novae coincide with our search fields, while we cover only 6.5\,\% of the entire sky. The third case might be too far south ($\delta = -43\degr$) to be reached by Far Eastern ancient observers. This arises an interesting question: Could this be a first hint for the existence of a new class of long-period recurrent novae whose eruptions repeat at time scales of centuries or millennia, in contrast to the hitherto known cases erupting at cadences of decades? We admit, the statistics are not significant and the effect could be accidental coincidence. Nevertheless, it might be worth to be mentioned. 

\begin{table}
	\centering
	\caption{Total number N of CVs of different subtypes listed in the VSX, AAVSO (retrieved November 2019, column 2) and their expected number within the total area of 2698.9 square degrees (6.54\,\% of the entire celestial sphere, column 3). For comparison, we give the numbers of our candidates, as listed in Tab.~\ref{tab:tab1} (column 4), the excess or deficit, compared to the expected number (column 5), and their mean errors (column 6, both in percent). }
	\label{tab:Tab5}
	\begin{tabular}{lrrrrr} 
		\hline
    CV            &N        & N       & N   & \multicolumn{2}{c}{excess or deficit} \\
subtype & total & expected  &shortlist  & \%   & $\pm$ error $\sigma$ (\% )  \\ \hline
UG      &  748   & 48.9     & 43      & $-12$   & 15 \\
NL      &  222   & 14.5     & 18      & $+24$   & 24 \\
AM      &  119   &  7.8     & 15      & $+92$   & 26 \\
DQ      &   59   &  3.9     &  8      & $+105$  & 35 \\
CV      &  305   & 19.9     & 11      & $-45$   & 30 \\
Nova    &  136   &  9.0     &  9      &  0      & 33 \\
LMXB    &   66   &  4.3     &  3      & $-30$   & 58 \\ \hline
Total:        &1655         &108.2             &107              &-1       &10  \\ \hline
	\end{tabular}
\end{table}

 Summarizing, the search in the VSX catalogue results in a representative subset of CV types. It appears to be a reliable source worthwhile to study.

 \section{Discussion of individual ancient events} \label{chap:discAncEv}
 In HVP2020 (Table 3), we had identified the stars or asterisms given in the texts with their modern names and defined our field of the CV search next to these positions according to the old texts. Here we give more details for each event discussing possible identifications with modern counterparts. First, we mention, in chronological order, those cases which require more extended explanations, either on details in understanding old texts, or on properties and references concerning possible identifications. Finally, we present in Table 3 shorter remarks on the remaining events studied here, also in chronological order.  

\textbf{In $-203$ (204 BCE)}, there is a `fuzzy star' reported from China which appeared close to the bright star Arcturus. There is no further information such as color or vanishing behavior. The day of the month is also not given (so the influence of the moon is not considerable) and the position is close to the galactic pole. Our shortlist of nova candidates consists of three stars. The classical nova Com 1877 (AB Boo) is only $2\fdg5$ away from Arcturus and was reported to reach $4.5$~mag in peak, as reported by \citet{schwab1901} 23 years after eruption date. Until now, no CV candidate at its position was identified casting some doubt on its existence. However, the dwarf nova RX J$1404.4+1723$ and the nova-like SDSS~J$143209.78+191403.5$ are other possible identifications. 

The event \textbf{in 64~CE}, also registered in Stephenson's (1976) lists of potential nova and supernova candidates, is described as seen next to $\eta$ Vir and `a guest star with bright vapors 2~chi or so long'. As two Chinese `chi' are almost 2\degr\ this appearance could have been caused by the atmosphere only if the appearance has been extraordinarily bright (at least as bright as Venus). For this event the brightest CV in the field of search is the AM Her type binary V379 Vir which turned out to be a CV with a sub stellar L-type donor \citep{stelzer2017}. It could be compared to the well-known nova V1500 Cyg which is also an AM Her type polar, showing in `quiescence' rather strong variability in V, ranging from $17$~mag to 21~mag \citet[Tab. 6.1]{Warner1995}. Therefore, V1500 Cyg has shown a total amplitude of the nova eruption $>19$~mag. Indeed, V379 Vir could have reached during eruption a similar brightness as Venus if it behaved like V1500 Cyg.

A total of 13 possible identifications were found for \textbf{event 70 CE}, among them also the doubtful Nova Leo 1612, that never has been identified with a modern counterpart, like the case AB Boo mentioned above. It is remarkable that these both doubtful historical novae coincide in some of our search areas, as discussed in Section 4. For this event, only the constellation is given (Xuanyuan, extending from the front feet of Leo to the feet of Ursa Major) and we selected it because of the given duration. 

Among the 10 candidates \textbf{for 668 CE} there is one case of special interest: V392 Per has been a poorly studied faint CV, detected in Sonneberg observatory by \citet{richter1970}, named S1065. \citet{liu+hu2000} presented a spectrum, dominated by H$\alpha$ emission as well as HeII\,4686, a line often present in nova remnants. The AAVSO archival light curves shows between 2012 and 2016 dwarf nova like outburst activity, with variation between 14~mag and 17~mag and it was observed in 2018, April 29th at 6.0 magnitude, with subsequent rapid decline, a typical nova eruption. The historical nova of naked eye visibility appeared in the constellation of Auriga (Milky Way) which was only visible in twilight those days. Thus, the phenomenon must have been much brighter than 6~mag. However, as the rise of the nova in 2018 was not observed, it could have been even brighter at maximum light and a coincidence with a possible nova in 668~CE is not completely excluded. We believe that this might be an excellent candidate for a recurrent nova whose earlier eruption was detected by ancient Far Eastern observers.  

For the \textbf{event 1437 CE} \citet{shara2017_nov1437} suggested a CV east of $\zeta$~Sco to have caused the event because they had identified a faint nebula as nova shell of this CV due to a proper motion analysis. The kinematic age of this nova shell-CV pair apparently dates the classical nova in the 15th century. However, as discussed by \citet{hoffmann2019} this CV+shell pair does not fit the historical position properly. In this case, the text reports a position `half a chi' separated from a certain star towards north and the shell is located east of this star. That is why, we insist searching in the area between the two given stars as suggested by all authors of coordinate lists of the last decades \citep[Fig.~6]{hoffmann2019} and will, then, decide with a look on the map whether or not a certain star is more likely than others because of its proximity to the $0\fdg5$ next to $\zeta$ Sco. This way, we found only one possible candidate, V1101 Sco, an interesting low-mass X-ray binary. However, looking at the map (Fig.~\ref{fig:event1437}) this valid candidate from the physical point of view must be dropped because it does not at all fit the described position between the 2nd and the 3rd star of asterism Wei (Sco)\footnote{There are three Chinese asterisms `Wei'. Here the one in the IAU-constellation of Scorpius is meant.}.

\textbf{1592 CE} is one of the most enigmatic ancient events in our sample: There are three coordinates given in the list of potential novae and supernovae by \citep{stephenson} and \citet[p.\,46-49]{steph77}. Apparently, Stephenson interprets even three different events while our list (in HVP2020) uses two fields of search. As we have shown in HVP2020 and in \citet{hoffmann2019} it would be wrong to consider the point coordinates of 20th century authors without error bars and in most cases, it is even misleading to give point coordinates at all because the area of a given constellations cannot be reduced to point. With appropriate error bars, the two coordinates of Stephenson's No.\,73 and No.\,74 overlap. As, additionally, the two guest stars are reported simultaneously (one between Nov. 30th 1592 and March 28th 1593, the other from Dec. 4th 1592 to March 4th 1593), we do not expect to learn different things about the evolution of binary systems if we match them to certain CVs: The historical novae would be of same age. That is why, we defined our field of search in a way covering both positions within the Chinese constellation of Wangliang. 

However, the duration of the $3-4$ months of visibility suggests a rather slow nova, the visibility in Cassiopeia (fainter clouds of the Milky Way) suggests a bright one and, thus, a possible postnova counterpart should be a bright one. Therefore, in our short list of candidate stars, the nova-like V592 Cas with $V\sim12.8 \pm0.4$~mag seems to be the most likely candidate for this event if it was caused by a classical nova eruption. In modern novae normally the decline rate is defined as $t_3$, the total time between maximum epoch and the passage of brightness 3~mag fainter than at maximum. The extreme example is RR~Tel 1949 with $t_3 = 2000$~days, but there are other slowly declining novae as V723 Cas 1995 with $t_3 = 299$~days or V849 Oph 1919 with $t_3 = 270$~days \citep{strope2010}. 

The other record from 1592 reports a guest star about three `cun' ($\sim0\fdg3$) away from the third star from the east in the constellation Tiancang. In our field of search around the `star in Tiancang' ($\zeta$ Ceti) we had originally chosen a radius of $1\degr$ around the star because the position is given as only 0\fdg3 away. In this small area no CV has been found. The duration of the historical appearance is reported from Nov. 23rd 1592 to Feb. 20th 1594: a total of 15 months! This is only comparable to the above-mentioned target RR~Tel, the slowest known modern classical nova, or perhaps it could refer to a supernova. However, this would imply a very strange coincidence, three or four galactic supernovae within 32 years because the given year (1592) is between the two well-known historical supernovae observed by Tycho Brahe (1572) and Joh. Kepler (1604). The dates as given in the original Korean counting of years are displayed in our Table~\ref{tab:tab2} which compares these four records. Looking at the dates in this table it is striking that the date of the SN~1572 is the 5th year of a certain reign while 1592 is the 25th of the same reign. Could it be possible that somebody (a later chronicler) misread the second entry of an enumerated list of records from the 5th year as part of the year number?

\begin{table*}
	\centering
	\caption{The strange Korean records of possible novae in 1592.}
	\label{tab:tab2}
	\begin{tabular}{lcccc} 
		\hline
		Record No. & 1 & 2 & 3 & 4 \\
					& SN 1572 		& 1592 north & 1592 south & SN 1604 \\
		\hline
	Starting date	& 5th year, 	& 25th year, & 25th year, & 37th year \\
					&10th month, 	&10th month, & 10th month, & 9th month \\
					& day [jiayin$=51$]	&day [guichou$=50$] &day [bingwu$=43$]& day [wuchen$=5$]\\
	asterism		& Cexing 		& Wangliang	& Tiancang	& Tianjiang \\
					&($\gamma$ Cas) &($\alpha,\beta\dots$ Cas) &($\zeta$ Cet) & ($\theta$ Oph) \\
	duration		& China: 15-16 months		&			&				& Europe, China: 12 months\\
					& Korea:  no record		&Korea: $3-4$ months	&Korea:	 15 months		& Korea: $\sim5$ months\\
	References: 	&  & & & \\
	Xu et al., 2000 &p. 143 &p. 144 &p. 144 & p. 145\\
		\hline
	\end{tabular}
\end{table*}

 Looking at the asterisms given in the Table~\ref{tab:tab2}, it also seems as if somebody (a later chronicler?) has mixed up information from the two collections of supernova records: The preserved texts are not the original reports of astronomers. A later chronicler summing up the history of a dynasty after a king's death could well re-write an observation next to the single star asterism Cexing ($\gamma$ Cas) as (of course truly also) next to the neighbouring asterism Wangliang (containing $\alpha$, $\beta$, and $\kappa$ Cas). This could explain, the missing observational reports of the SN 1572 in Korea: The record No.~2 in the table would then be the Korean observation of Tycho's supernova (the year given corruptly as 25th instead of 5th year of a certain king's reign).\footnote{The different names of first sighting do not argue against this interpretation: First, names of days usually differ in Korea from those in China. Second, the day of a sighting is influenced by weather and other constraints.} The record no.~3 could, then, be a confusion of information from both supernovae: the date should be 5th year because the duration fits the SN~1572, whilst the constellation should be Tianjiang instead of Tiancang (pronounced similarly) fitting the SN~1604. Thus, the record(s) preserved from Korea, 1592, could -- if we take them seriously -- report an observation of one or two novae or supernovae but they could also be a result of errors in the copying tradition. As the position of SN~1572 fits the description for one of the events perfectly we give priority to the interpretation as writing errors.

 Speaking of supernova interpretations, the idea of identifying the sighting of 1592 with the explosion which caused the supernova remnant Cas~A might come up \citep{brosche1967,chu1968}. We consider this as hardly possible because $(i)$ the distance of Cas~A to the closest star of the asterism of Wangliang is $\sim6\degr$ and, thus, this object is located in another Chinese constellation which is called the Ascending Serpent on the Suzhou map \citep[p.\,20, no.\,65]{rufus} or the Flying Serpent since the Su Song map of the 11th century: We sketched the scenario in our Fig.~\ref{fig:event1592W}. $(ii)$ The age of the SNR Cas~A has been roughly hundred years younger: \citet{fesen2006} applied various methods to determine the age and derives three possible values ($1662\pm27$, $1672\pm18$ and $1681\pm19$) leading to an earliest possible explosion date of 1635 and a bigger likelihood for the second half of the 17th century.

\textbf{The 1690 CE} event is reported in two ways \citep[p.\,146]{xu2000}: One giving Chinese coordinates by instrumental measurements (at a time after invention of the telescope) and one placing it traditionally next to a single star. Our selection criterion to put it in our shortlist was the position given next to a single star: east of the third star of Ji.

Why not taking the coordinates which appeal us due to their higher accuracy: The result of the instrumental measurement is reported as (RA,DE)$_{1690_\star} =$(3~du, 18 fen in WEI, $-34$ du, 20 fen).\footnote{1~du $= 360/365.25=0\fdg99$, one fen is a minute of arc.} That means, the declination is $-34\fdg\bar{3}$ and the right ascension is computed as `right ascension of the determinative star of the lunar mansion Wei plus $3\degr 18\arcmin$, implying that the transient was at RA$=250\fdg0$ in 1690. The Chinese system of counting right ascensions started from the determinative star of each lunar mansion eastwards and restarted the counter at the determinative star of the subsequent lunar mansion. The constellation and lunar mansion of JI is \textit{east} of the neighbouring constellations and lunar mansion WEI. Hence, an object east of any star in Ji would \textit{not} be given in `du in WEI' but `du in JI'. Thus, the given tuple must be corruptly preserved. In principle, also the single star description could be wrong. Then, it would not be meant the `3rd star of Ji' but the `3rd star of Wei' and the coordinate would be correct. However, this is unlikely because: $i)$ There are two independent records reporting the 3rd star of Ji and only one mentioning the coordinates. Thus, the chance for a writing error in the coordinates is higher than in the putatively imprecise other position. $ii)$ The 3rd star of Wei is roughly at the same right ascension as the 1st star of Wei and further south than the given declination. In contrast, the 3rd star of Ji had the coordinates (RA,DE)$_{1690,Ji^3} =(270\fdg9,-34\fdg5)$ and the determinative star of Ji ($\gamma$ Sgr) was at RA$_{1690,Ji^1}=266\fdg9$ which leads to a perfectly fitting RA of the transient (RA,DE)$_{1690_\star} =(270,-34.3)$. Taking into account an error bar for the instrumental measurement of $\sim1\degr$ and the possibly old calibrations it is well possible that the transient appeared not exactly at the 3rd star of Ji but slightly east of it as reported. Thus, we consider the coordinates preserved corruptly and the position `east of the 3rd star of Ji' as given correctly.

 The designation `3rd star of Ji' refers to a position east of $\epsilon$ Sgr. The transient lasted only 2~days and the area becomes visible roughly in the south during darker dusk. There was no moon in the evening sky (illuminated $\sim10$\,\% this night and waning) and Venus as evening star stood next to Saturn. The area, where the guest star appeared was only $\sim20\degr$ above the horizon and set around 10~pm (4 hours after sunset). During these hours the sky was dark; the nova could have any visual magnitude but due to its proximity to the horizon and the bright Milky Way, 3~mag and brighter are more likely. As the colour of the transient is given in the text (yellow), it likely had a brightness $<2$~mag (because otherwise the human eye would not be able to distinguish colours)\footnote{The detection limit for colours of stars is between 2 and 3~mag.}. As the historical event lasted only two days, it suggests having been caused by a fast nova with, thus, large amplitude. Among our candidates for this event, there are two fast novae V909~Sgr and V2572~Sgr with peak brightness 6.8 and 6.5~mag, respectively. This is slightly below the threshold of naked eye observation but modern recurrent novae show considerable variations among the different nova eruptions observed, for instance T~Pyx \citep{mayall1967,schaefer2010} (peak brightness varies by 1~mag) or U~Sco (peak brightness varies by 2 to 3~mag, see AAVSO, light curve archive). The slow nova V4745 Sgr seems to be a less probable identification (too faint with 7.3~mag peak brightness); the same is probably valid for the three dwarf novae in this field. 

A similar analysis as shown above has been performed for the other events in Tab.~\ref{tab:tab1} but we present only a summary of the results in the Table~\ref{tab:novaCandDiscuss}.  

Finally, three events in our shortlist (HVP2020, their Table~3) are missing here in Table~\ref{tab:tab1}: $-103$, $-4$, and $+1497$. For two of them ($-103$ and $+1497$) no CVs at all have been found although the positioning of the historical observations is rather precise. For the third event $-4$ (5 BCE) we found a total of 11 CVs but all of them are too faint: As they failed our criteria described in Section 2 they are incompatible with classical nova eruptions of naked-eye visibility. 

\begin{table*}
	\centering
	\caption{Additional Information on our events in the historical text (column~2) and comments on their potential meaning (column~3) for the identification of modern counterparts (column~4). Last column gives the number (no.) of remaining candidates.}
	\label{tab:novaCandDiscuss} 
	\footnotesize
	\begin{tabular}{rp{.14\textwidth}p{.29\textwidth}p{.36\textwidth}r} 
year
  &Information 
  &Conclusions 
    from the ancient
         description 
  &Possible identifications & no. \\\hline
$-203$ 
	&  see text
	&
	& 3 remaining candidates, brightest one: AB Boo 
	& 3\\ 
$-103$ 
	&  see text
	&
	& no CVs in the field 
	& --\\
$-47$
	&As large as a melon, bluish-white
	&Negative magnitude (brighter than Venus, i.\,e. $-4$ mag or brighter) 
	&The brightest two CVs in the field have normal levels around 18 mag and reach 15 mag only in DN outbursts. They can unlikely cause such a bright nova. 
	See also a discussion on this event in \citet{hoffmann2019}.
	& 0\\  
$-4$ 
	&see text 
	&
	& no candidate found. 
	& 0\\ 
$+64$ 
	&see text 
	&
	& only remaining candidate V0379 Vir is too faint.
	& 0\\ 
70 
	&see text 
	&
	& 13 equally valid candidates remaining. 
	& 13\\ 
101
	&Small star next to $\alpha$ Lyn (3.14 mag), bluish-yellow
	&\multirow{2}{.29\textwidth}{A little bit fainter than 3 mag but not too faint to be visible next to a 3 mag star: $4-5$ mag peak brightness}
	&\multirow{2}[5]{.39\textwidth}{Two likely candidates: BK Lyn (already suggested as counterpart by \citet{pat2013}; see also \citet{hoffmann2019}) and SDSS J100658.40+233724.4. The third and faintest target YZ LMi is an AM CVn-type binary, consisting only of He. It has to be discarded as a candidate.}
	& \\
or 101
	&Small star next to $\zeta$ Leo (3.4 mag), bluish-yellow
	&
	&\multirow{2}[2]{.39\textwidth}{ .}
	&2
\\
\multicolumn{4}{c}{.}\\
329
	&No information
	&
	&`Trepassed against Beidou' means approaches the N. Dipper within 1 degree. (!)Four stars within the Dipper's bowl: asterism Tian Li (Celestial Judges), see Fig.~\ref{fig:event329}. $\rightarrow$
SDSS J113215.50+624900.4 would be considered in The Judges, not possible as candidate. GP CVn and CV Boo are $4-5\degr$ away from the line of Beidou, DW UMa and SBS 1316+577A $2-3\degr$, which makes them less likely.
	& 5(+4)\\
	&&&5 candidates remaining, \\
	&&&$+4$ less likely ones.\\
641	&No information
	&during dusk in the west, sun approaching (visible only in twilight)
	&Post-nova brightness: peak 2 mag $+13$ mag (amplitude) $=15$ mag. IR Com is too faint (normally $V\sim16.2\dots17.8$~mag).
   Most likely candidate: the NL/VY star SDSS J122405.58+184102.7
	& 1\\
667	&No information
	&In twilight, close to horizon $\rightarrow$ bright, $\sim2$ mag
	&Post-nova brightness $\sim15$ mag. 3 remaining candidates, among them V518 Per, a low-mass X-ray binary 
	&3\\ 
668	&No information
	&Likely in Milky Way (Aur), in twilight $\rightarrow$ bright, det.limit $\sim2$ mag
	&Post-nova brightness $\sim15$ mag (see text!), 
	6 candidates remaining 
	&6\\
683 &No information 
	&possibly next to Milky Way (Aur), $\sim3$ mag(?)
	&Post-nova brightness $\sim16$ mag, 
	3 stars possibly too faint. 1 remaining.
	&1(+3)\\
722
	&No information
	&In Milky Way (Cas): 2.5 mag limit. If decline $3-4$ mag within 5 day: very fast nova.
	&A total of 9 candidates; among them Nova V723 Cas 1995 (modern peak: 7.1 mag, if real peak was missed: $\sim6$, if historical peak 2 mag brighter: 4 mag. unlikely in Milky Way but not impossible). Identification with HT Cas was suggested by \citet{duerbeck1993}.  
	& 9\\
840
	&No information
	&In twilight (dawn) $\rightarrow$ 2 mag limit
	&5 candidates, among them the intermediate polar and dwarf nova IP Peg
	& 5\\
891
	&No information
	&
	&2 candidates, one of them is the recurrent nova U~Sco
	& 2\\
1175 &As small as Mars, dense rays
	&\multirow{2}{.29\textwidth}{Mars: 1.28 mag$\pm 0.3$ mag (because Mars next to Spica: 0.95 mag and Jupiter: $-1.79$ mag; Mars is red, thus brightness likely underestimated).}
	&Expected post-nova brightness: $\sim15$ mag. If decline $>4$ mag within 5 day: very fast nova (extremely fast!). The two candidates in our shortlist of candidates seem to be too faint. &0\\
	&&&0 candidates remaining. &\\
1430
	&Bluish-black, as large as crossbow pellet
	&`large' (extended) and visible beyond the Milky Way means bright, but of dark colour. If decline $>5$ mag within 26 day: very fast nova. 
	&At least 2 mag peak brightness (because colour visible), post-nova brightness $\sim15$ mag. The DQ Her type star BG CMi is more likely than the remaining two CVs in this field because both seem to be too faint.
	& 1(+1)\\
1431
	&Yellowish white, shiny bright, as large as a crossbow pellet
	&Appeared close to constellation Orion+Sirius, close to Milky Way, thus magnitude at least as the brightest star Sirius: $-1.45$ mag. If decline $>6$ mag within 15 day: very fast nova. 
	&Bright post-nova $=13$ mag expected. Among 10 stars in Tab.~\ref{tab:tab1} the fast nova KT Eri (peak 5.4 mag) would be a promising recurrent nova candidate in case of variability of peak brightness (eclipsing Z And type); AQ Eri and BF Eri are possible. 
	& 9\\
	\end{tabular}
\end{table*} %
\begin{table*}
	\centering
	\contcaption{ }
	\footnotesize
	\begin{tabular}{rp{.14\textwidth}p{.29\textwidth}p{.36\textwidth}r} 
		\hline
year
  &Information 
  &Conclusions 
    from the ancient
         description 
  &Possible identifications & no. \\
	\hline
$1437$ 
	&see text 
	& 
	& 1 candidate: V1101 Sco (low-mass X-ray binary); must be dropped (see Fig.~\ref{fig:event1437}). 
	& 0(1) \\ 
1461
	&A star as white as powder
	&
	&3 candidates have similar likelihood, among them especially V380 Oph (NL/VY) & 3\\
$1497$ 
	&  see text
	&
	& no CVs in the field & 0\\
$1592$ 
	&see text 
	&
	& possibly a corrupt text (no nova). & [no]\\ 
1661
	&As large as Saturn (0.5 mag)
	&
	&Post-nova: 0.5 mag peak $+ 13$mag $= 13.5$ mag. Our two candidates are much fainter, 0 remaining. &0\\
$1690$ 
	&see text 
	&
	& 3 candidates remaining. &3\\ 
		\hline
	\end{tabular}
\end{table*}

 \section{Discussion of the resulting list of candidates} 
 The resulting list of this evaluation is displayed in Tab.~\ref{tab:tab1} reduced of the stars which we labeled as `too faint' in column~12 for historical reasons. 

 \subsection{Remarks on the statistics of the final list}
 In this section, we refer only to those 80 targets (out of the 107) which have been selected according to the criteria based on the information in the ancient text as described in Section~\ref{chap:discAncEv}, including Tab.~\ref{tab:tab2} and Tab.~\ref{tab:novaCandDiscuss}. Thus, we do not consider the stars marked as `too faint' in the last column of Table~\ref{tab:tab1} anymore.
 
 The combination of Tab.~\ref{tab:tab1} with the additional information displayed in Tab.~\ref{tab:novaCandDiscuss} results in a much shorter list of real candidates to be able to brighten up to naked-eye visible magnitudes. Some of the theoretically possible CVs are not bright enough for the particular event which must have been extraordinarily bright (e.\,g. in case of event $-47$) and, thus, some of the events provide no appropriate CV or LMXB which could be an appropriate counterpart for the historical event (see Tab.~\ref{tab:Zusf}). For the best determined and interesting event 1437 \citep{shara2017_nov1437}, we could suggest  V1101 Sco from the astrophysical point of view. However, this object does not fit the position (see Fig.~\ref{fig:event1437}). Summarizing, we cannot suggest a CV or X-ray binary to have caused the historical Far Eastern sightings in $-103$, $-47$, $-4$, $+64$, $1175$, $1437$, $1497$, and $1661$. For 1592 we found eight candidates which are bright enough to be able to cause a naked eye nova but it appears also likely that these Korean records are corrupt and do not report a nova at all (but the supernovae in 1572 and 1604). Hence, for 9 out of 24 events in our shortlist (more than a third) we did not yet identify any hint that they are records of classical novae. We will present further investigation in a later paper. 
 
 In contrast, for $641$ and $1437$ we found only one candidate, and for the sighting in $683$ one out of four candidates is more likely than the others. So for the events $641$ and $683$ we suggest SDSS J122405.58+184102.7 and ASASSN-14gy, respectively. Similarly, for the event in $1430$ one out two (BG CMi) and for the appearance in $1431$ two out of nine (and BF Eri even more likely than AQ Eri) are more promising than the others. Thus, further research on these objects will likely turn out the modern counterpart of the historical sighting. 
 
  \subsection{Remark on the durations of the historical events}
Table~\ref{tab:Tab4} also lists the durations of 15 ancient events, for which some information on their duration is available, ranging from 2 days to several months (see also Table~3 in HVP2020). It is well known that rapidly declining novae have larger amplitudes than slow ones \citep[chapter 5.2]{Warner1995}. Therefore, in average we expect brighter stellar remnants for the longer lasting events than for those with short duration, as considered in our discussion of individual targets in Section~3. However, we should recall that a duration given in the context of a chronicle as in the example of guest stars (and not in an astronomical diary) could also refer to the duration of its astrological significance instead of the real visibility \citep{hoffmann2019}. Only in some cases, the date of vanishing is given implying a real observation. 

\begin{figure}
    \caption{Upper panel: Histogram of the decline time $t_3$ of modern novae from \citet{strope2010}. Lower panel: Histogram of visibility durations of ancient guest star observations. Binning 10 days. Both distributions coincide within the errors.}
    \label{fig:histogr}
	\includegraphics[width=\columnwidth]{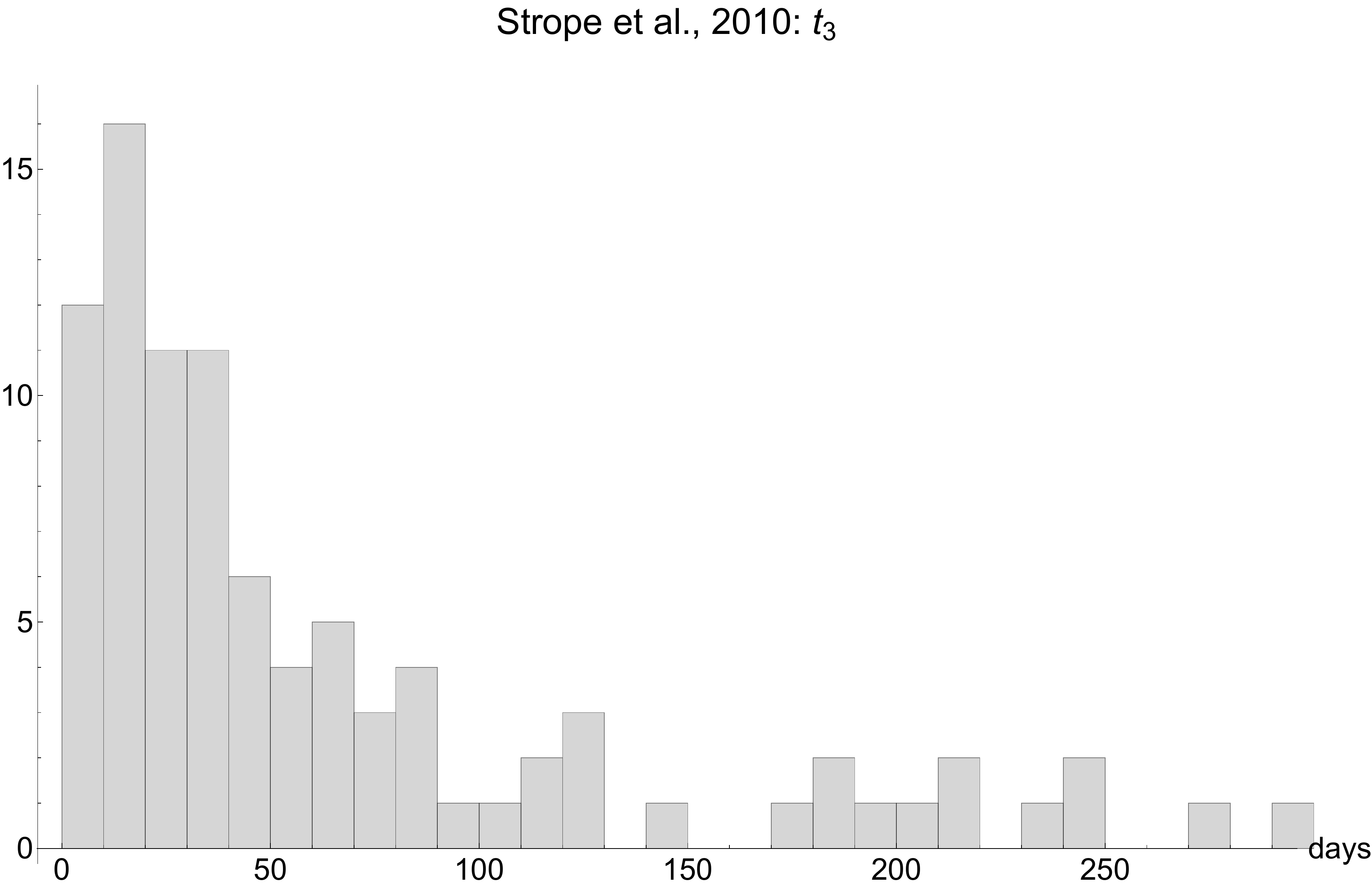} \\
	\includegraphics[width=\columnwidth]{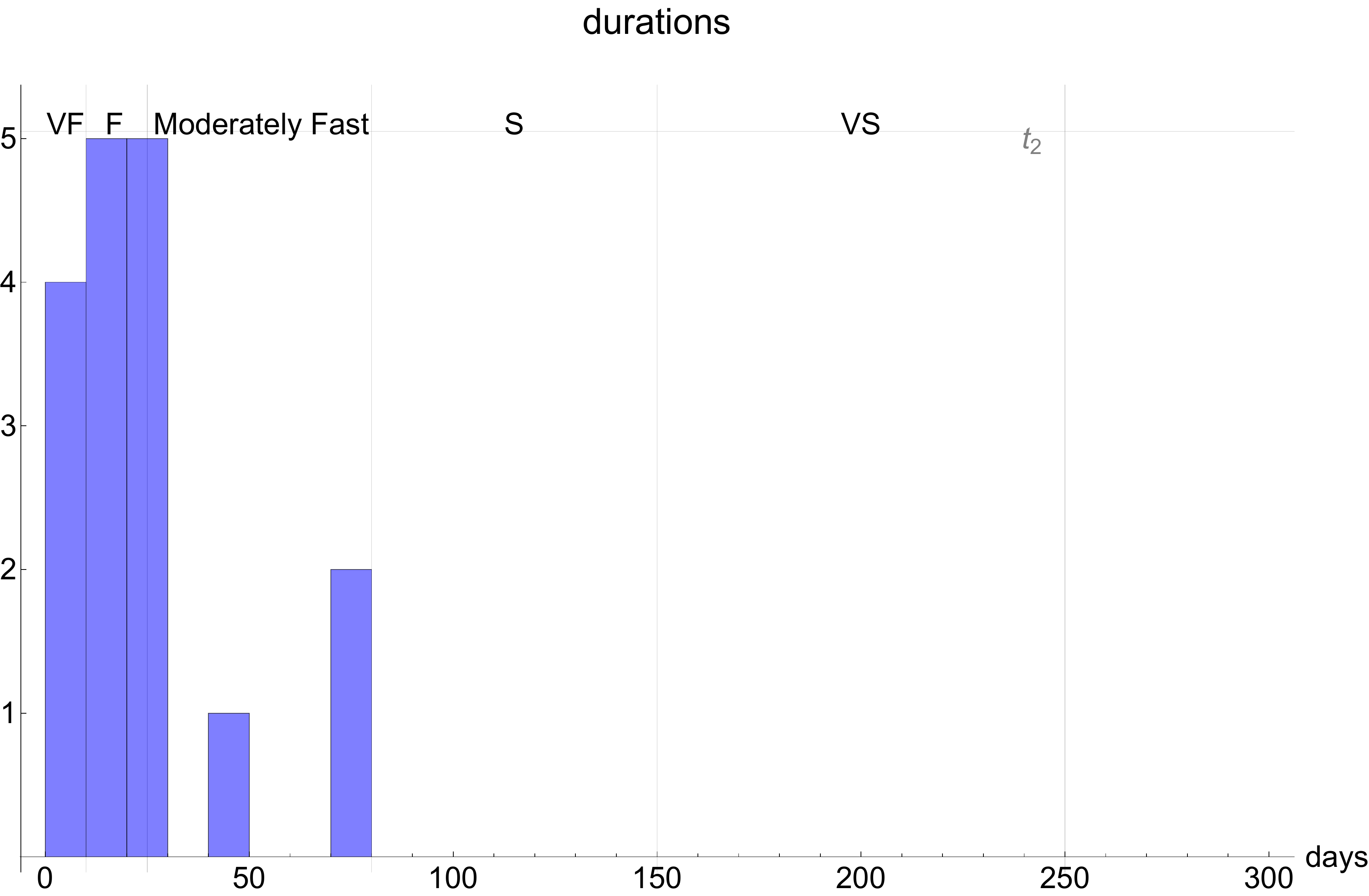}
\end{figure}

In addition, we can compare the distribution of the ancient durations with the $t_3$ decline speed of modern novae (reminding that some of these durations likely witness a decline of more than 3 mag, cf. Tab.~\ref{tab:novaCandDiscuss}). Fig.~\ref{fig:histogr} shows this comparison. Both distributions coincide within the errors as expected: if the mean maximum brightness of ancient classical novae is of the order $2-3$~mag, $t_3$ will refer to the epoch when the fading nova has reached $5-6$~mag, i.\,e. the limiting magnitude of naked eye observations. On the other hand, this coincidence can be taken as a possible hint that most of the targets presented in this study could, indeed, refer to classical novae and further investigations appear worthwhile.

\section{Summary, Conclusion, and Outlook}
HVP2020 had selected 24 especially promising guest star events among several hundred records available from ancient Far Eastern astronomers as a representative subset of historical records for possible use in modern astrophysics -- either because of well-defined positions or because of a duration given. This is considered to be a manageable amount but yet a representative subset of the 75 possible novae and supernovae in \citet{stephenson} enlarged by 9 events found in \citet{ho} and/ or \citet{xu2000} and newly located. One of the 24 events (the Korean records of 1592) turned out to be likely corrupt descriptions of something else, so we consider 23 promising historical nova observations. In this paper, we apply our suggested strategy to identify modern counterparts for this subset of historical records in case they really had been classical novae. This way, we hope to contribute to the questions on century-time scale of the evolution of close binaries and cataclysmic stars, e.\,g. \citep{pat2013,miszalski2016} and \citep[and subsequent of the `Life after Eruption' project]{tappert2012}. 

Our discussion took place in the following steps: 
 \begin{enumerate} 
  \item Re-read the historical record and define the area in the sky where the transient is reported (HVP2020). 
  \item Partition these areas with search circles for the VSX Search (HVP2020). 
  \item Find all CVs and X-ray binaries in the field with VSX Search and select those which are brighter than 18~mag (Tab.~\ref{tab:tab1}). 
  \item After ensuring that this subset is representative (Section~\ref{sec:stat}): re-consider the stars of Tab.~\ref{tab:tab1} with the conditions in the particular case which could shift the magnitude limit to a brighter one (e.\,g. was it only visible in twilight, did the Moon stand close by, or was it as 4~mag peak unlikely detected within the Milky Way?): Tab.~\ref{tab:novaCandDiscuss}.
  \item The result is a table of 80 naked eye candidates (Tab.~\ref{tab:tab1}, reduced by the `too faint' ones).
 \end{enumerate} 

After re-considering each case of the selected historical events in particular there are much less candidates remaining than found in the field and listed in Tab.~\ref{tab:tab1}. The totals are summarized in Tab.~\ref{tab:Zusf}. 

\begin{table}
	\centering
	\caption{Number of likely nova candidates}
	\label{tab:Zusf}
	\begin{tabular}{rr|rr|rr} 
		\hline
Event  	&N$_\star$ &Event  	&N$_\star$ &Event  	&N$_\star$\\
$-203$   &3 	&641    &1 		&1430        &1(+1)\\
$-103 $  &$-$  	&667    &3 		&1431        &9 \\ 
$-47 $   &0 	&668    &6 		&1437       	&0(1)\\
$-4 $  	 &0 	&683    &1(+3) 	&1461      	&3\\
64       &0 	&722    &9	 	&1497       	&$-$\\
70       &13	&840    &5 		&1592-1    	&[0] text corrupt?\\
101      &2 	&891    &2 		&1592-2     &[8] text corrupt?\\ 
329      &5(+4) &1175	&0		&1661       	&0\\
						& & &   &1690        	&3\\
		\hline
	\end{tabular}
\end{table}

 \textbf{Outlook:} For our selection, we applied clear physical criteria as described in HVP2020 and for this work, we selected a shortlist of records with more detailed information out of 185 records matching our criteria. Hence, the method developed in this paper should be applied to all other entries of the 185 entries and the more uncertain records with ambiguous descriptions are not yet considered in HVP2020 and Tab.~\ref{tab:tab1}. It could be well possible that in some of the remaining old records classical nova events are hidden. For instance: A possible nova or supernova record should not report any movement among the stars, the object should be steady. In contrast, one of the records of the SN~1006 from China reports `It passed through the east of [constellation] Kulou.' \citet[p.\,137]{xu2000} which means, with our physical criteria we would neglect it as supernova candidate. Thus, the strategy to continue this threat in research will be: 
\textit{First,} proceed with a broader study of the remaining $\sim160$ historical records with no reported movement and tail (but less precise position and no given duration) in the same manner as applied above and explained in HVP2020 and here, i.\,e.: Step 1: An honest positioning of the historical description in areas of search instead of guessing wrong point coordinates. Step 2: Search for promising candidates (of CVs and alternatives). Step 3: Follow-up observations of all strange and interesting objects in the field -- be they CV, LMXB to examine whether or not they could have caused novae and even supernova remnants or whatever object which would be able to cause an appearance as described. The method of this observation, of course, depends on the object: For gaseous shells the expansion rate and proper motions should be determined \citep{shara2007,shara2017_ATcnc_steph,shara2017_nov1437,schaefer2010}, for neutron stars the period, for CVs the spectra have to be checked for the presents and strength of the HeII-4686 line, for example. \textit{Second,} find further candidates for possible guest stars in old records although their terminology and wording could imply a movement. In such cases, the event should be ensured to be steady by other sources -- e.\,g. by observation from other cultures (might they be Babylonian, Greco-Roman, native American, Polynesian, Aboriginals or anywhere else in the world). 

That is why, we are currently also working on a database of historical positioning in order to provide a reliable source to look up potential historical events after finding an interesting astronomical object. This will improve the usability of historical data for astrophysicists not trained in dealing with them and enrich mankind's infrastructure of astronomical data. 

As already mentioned in Section~\ref{sec:CVlist}, Tab.~\ref{tab:tab1} displays only a shortlist of the most likely counterparts among currently known CVs and LMXBs. The results of search for alternative explanations of these historical will be the scope of our next paper(s). The VSX catalogue, used for the current purpose, is rather dynamic and adds new CVs nearly every day, most of them are very faint dwarf novae. Therefore, our shortlist reflects the situation at the end of 2019. However, all parameters of the search areas have been published by HVP2020, any interested researcher can repeat the same search in the VSX at any time in order to obtain actual results. 

Additionally, the magnitude limits used in this study could be modified and adapted towards newly found fainter sources. However, fainter limits would increase the number of chance coincidences (cf. VHT2019 for the limits and areas used in this study). On the other hand, it is possible that the VSX already contains the real counterpart of some of our ancient events, but it has been erroneously classified as another variability type, this way escaping our detection. Therefore, our presented results are only a first step towards a more systematic search with the goal of correct identification of at least some of the ancient Far Eastern records with classical novae (or, more generally spoken, to explain the sightings as nova or something else but this is not the scope of this paper). The further goal is to study their stellar remnants in more detail and to be able to derive conclusions concerning the postnova evolution millennia after the eruption. This task is not easy, as shown in the recent study by \citet{hoffmann2019}. Real evidence for one of, in average, three candidates per event listed in Tab.~\ref{tab:tab1} will only be achieved with laborious observational follow-up studies. They should comprise extended photometric long-term coverage in all possible time scales, between minutes and years, in order to know better the detailed behavior of each candidate. The LSST \citep{kessler2019} will help here, as soon as it comes into operation. This task should be complemented by multi-band spectroscopy, in order to identify emission lines typical of nova remnants (HeII 4686 is one of them), and to determine orbital periods and other binary properties for targets without present knowledge of these parameters. Another way to find better evidence could be the search for expanding nova shells around possible postnova candidates. This was already done in other contexts, not related to ancient Far Eastern guest stars \citep{himpel1942,Schmidtobreick2015,Schmidtobreick2015b} but, to our knowledge \citep{hoffmann2019}, not systematically starting from an ancient sky position except the single case of the Nova Sco~1437 \citep{shara2017_nov1437,sharaSearch1437} which is possibly not yet correctly identified \citep{hoffmann2019}. Finally, also other types of targets, which are not considered here (symbiotic stars of Z And type and super soft X-ray sources) could be investigated as possible counterparts, increasing even more the general difficulty to apply valid criteria for a unique and certain identification. Additionally, we of course have to check for alternatives to novae as explanation of the sightings because due to the lack of historical follow-up observations and the resulting disability to draw light curves we can never be sure that a certain event really refers to a nova. This further cross-check is already in preparation for publishing. Thus, we already are able to state that it will not contradict this study.   

\section*{Acknowledgements}
S.H. thanks the Free State of Thuringia for financing the project at the Friedrich Schiller University of Jena, Germany. N.V. acknowledges financial support from FONDECYT regularNo. 1170566 and from Centro de Astrofísica, Universidad de Valparaíso, Chile. In terms of Chinese language, we thank Jesse Chapman (University of California, Berkeley), Nalini Kirk (HU, Berlin, Germany), and David Pankenier (Lehigh University, Bethlehem, Pennsylvania) for useful hints. Thankfully we made use of the VSX variable star catalogue of the American Association of Variable star Observers (AAVSO) and of the SIMBAD data base (Strassbourg). Ralph Neuhäuser (AIU, Friedrich-Schiller-Universität Jena) had the initiative and idea to reconsider historical nova identifications including new nova candidates in a transdisciplinary project. We thank him to have brought us together.




\bibliographystyle{mnras}
\bibliography{cvCand} 


\newpage

\appendix
\definecolor{darkgreen}{rgb}{0,0.8,0} 
\newpage

\section{Online-Only Material}
Two maps for each of the 25 selected ancient events in 24 years which could refer to classical novae: The left map shows our search circle(s) and all CVs in the area, the right map shows only the selected candidates (Tab.~\ref{tab:tab1}, reduced by the `too faint' ones). All maps in equatorial coordinates (epoch 2000). CV marker: green \textcolor{darkgreen}{\rotatebox{90}{$\diamondsuit$}} while the cyan coloured \textcolor{cyan}{\rotatebox{90}{$\diamondsuit$}} indicates X-ray binaries. The CVs which are considered to be possible candidates are additionally highlighted by a green \textcolor{darkgreen}{$\star$} symbol. 

At the upper edge of the chart frame, the event is again written and its duration (in \textbf{d}ays or \textbf{m}onths) is given. At the right edge of the frame, the Chinese asterism is given in which the mentioned in the text. In case of the transient in 667, the position is given between three asterisms which are, therefore, highlighted in our chart. A legend next to the right frame gives the designations of the candidates displayed in the chart.  

The stick figure lines of the Chinese asterims are taken from Stellarium 0.19.2 (version by contributed by Karrie Berglund, Digitalis Education Solutions, Inc. based on Hong Kong Space Museum star maps), slightly varied in cases where the identified star was not in the Yale Bright Star Catalog (HR). This version contains a few more asterisms than the Suzhou map which is why our search circles are developed by always cross checking with the original historical map (as described in our earlier publications).

\begin{figure}
	\includegraphics[width=\columnwidth]{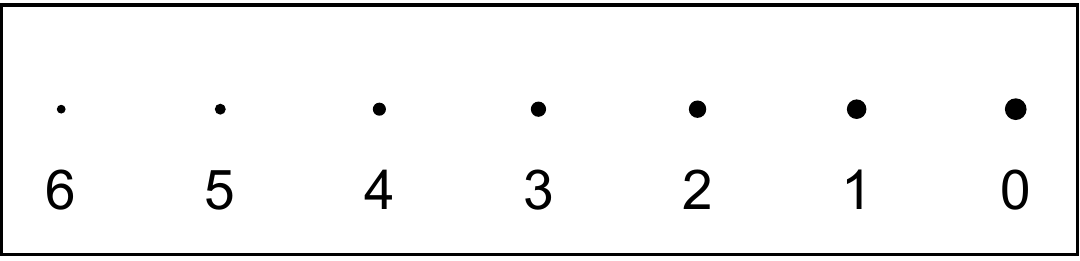}
    \caption{The point sizes of stars in the charts are scaled by magnitude.}
    \label{fig:magLegend}
\end{figure}

\begin{figure*}
	\includegraphics[width=\textwidth]{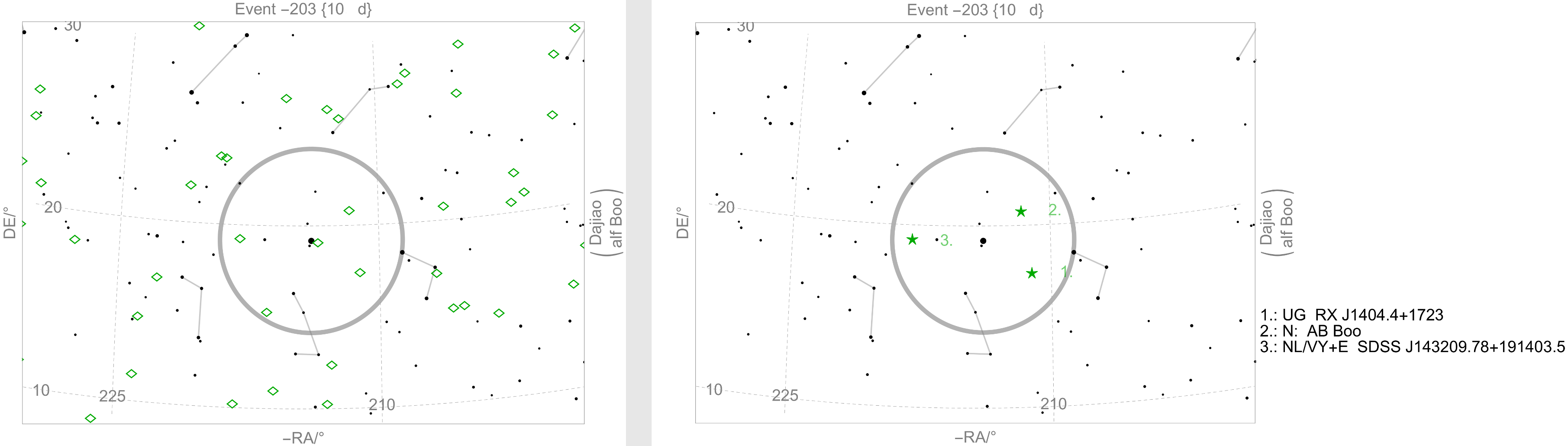}
    \caption{Event $-203$: The left chart displays all the CV in the area; the right chart displays the selected candidates.}
    \label{fig:event-203}
\end{figure*}

\begin{figure*}
	\includegraphics[width=\textwidth]{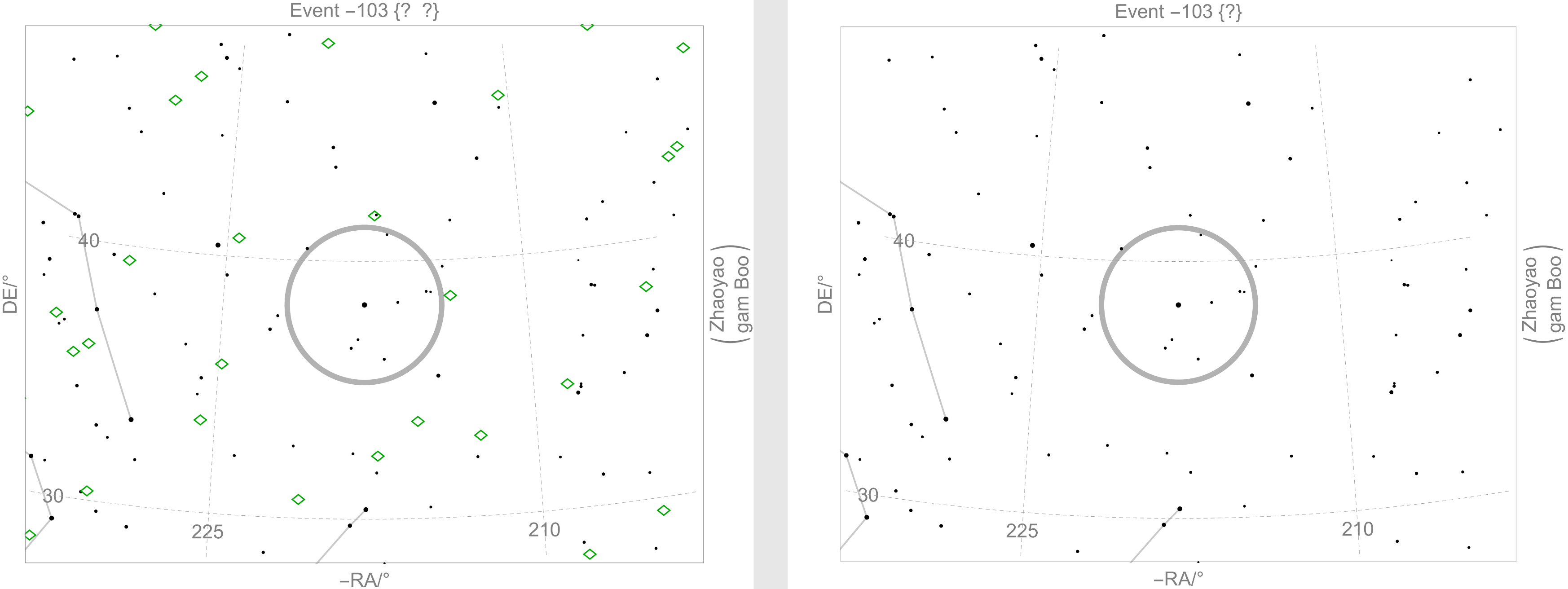}
    \caption{Event $-103$.}
    \label{fig:event-103}
\end{figure*}

\begin{figure*}
	\includegraphics[width=\textwidth]{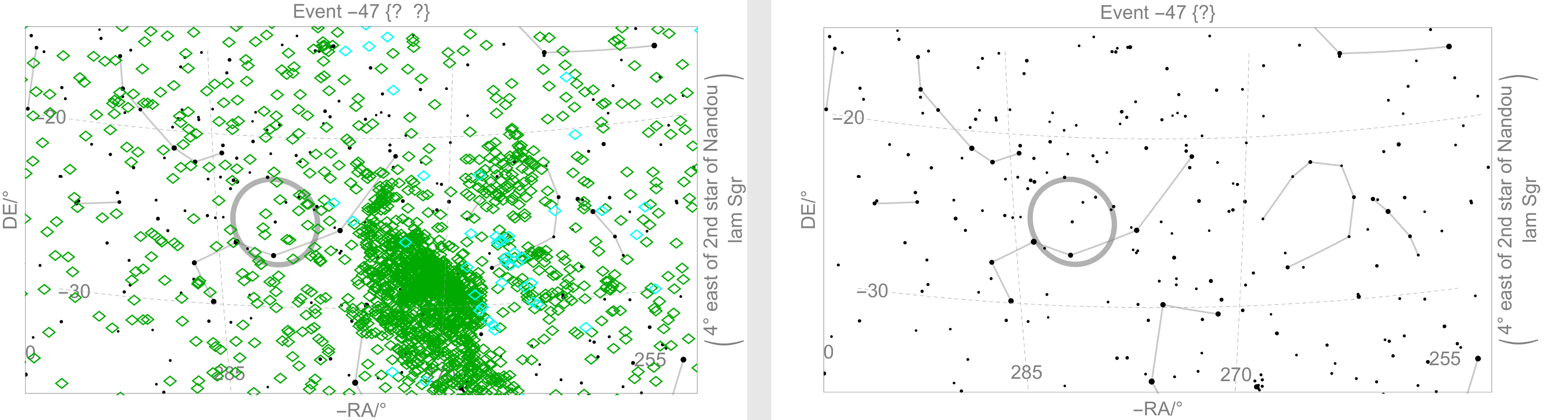}
    \caption{Event $-47$.}
    \label{fig:event-47}
\end{figure*}

\begin{figure*}
	\includegraphics[width=\textwidth]{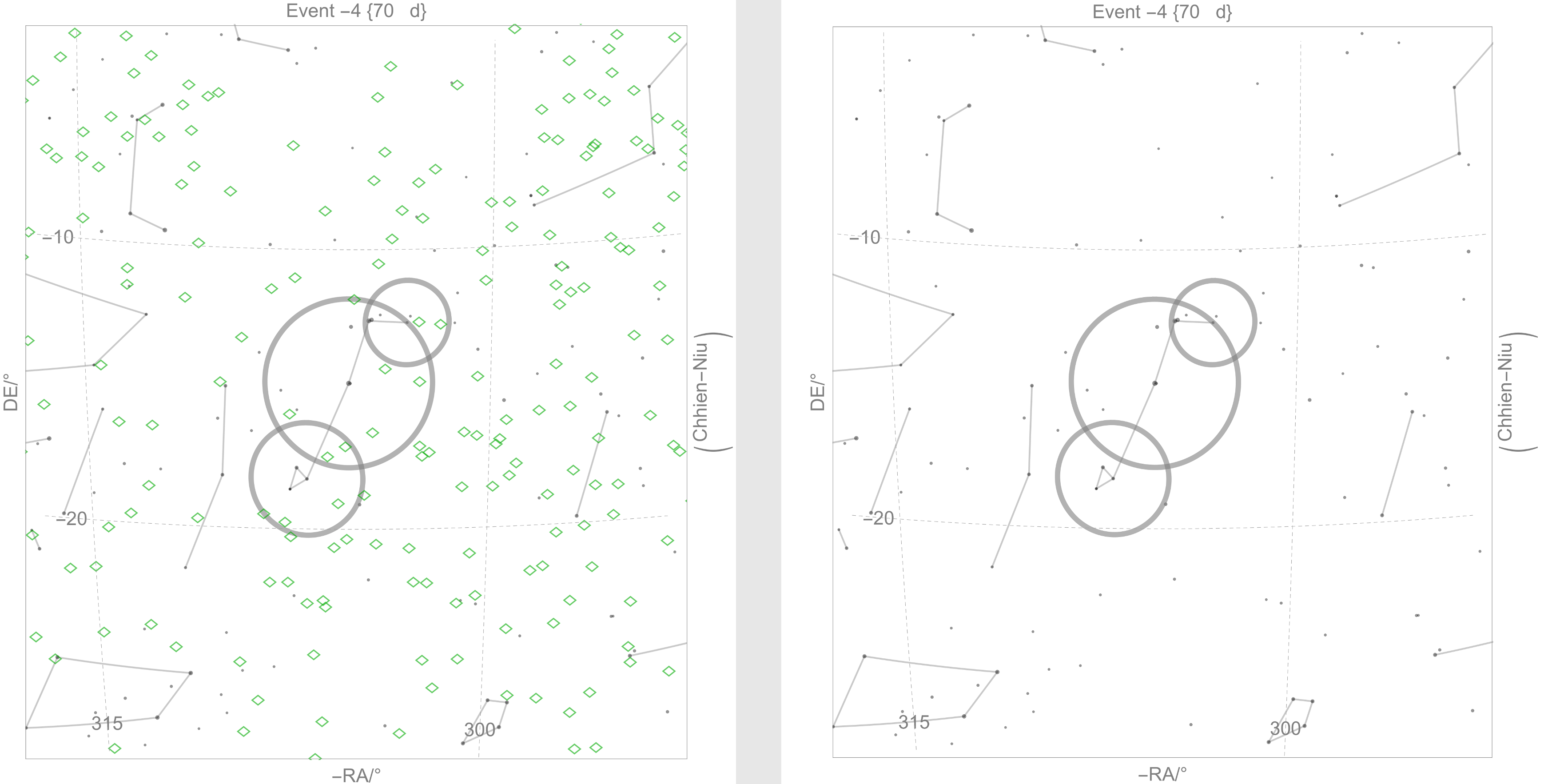}
    \caption{Event $-4$.}
    \label{fig:event-4}
\end{figure*}

\begin{figure*}
	\includegraphics[width=\textwidth]{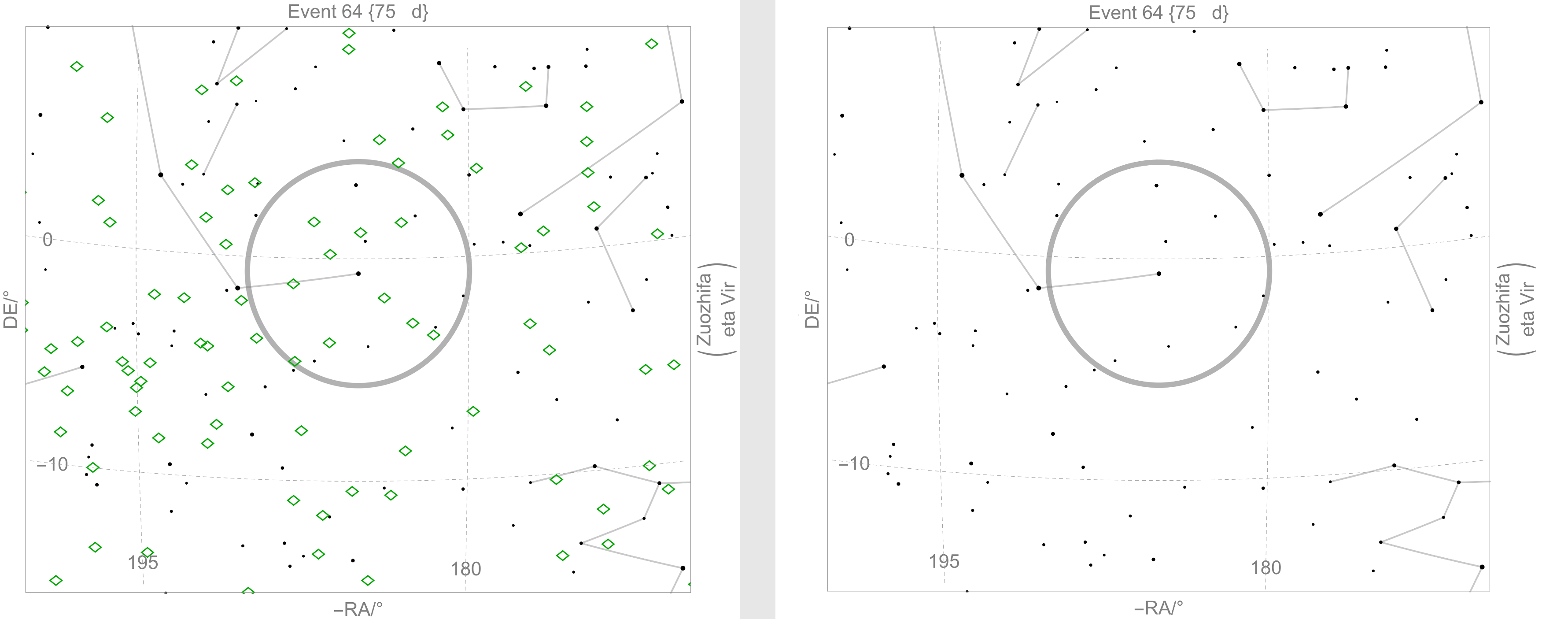}
    \caption{Event $+64$.}
    \label{fig:event64}
\end{figure*}

\begin{figure*}
	\includegraphics[width=\textwidth]{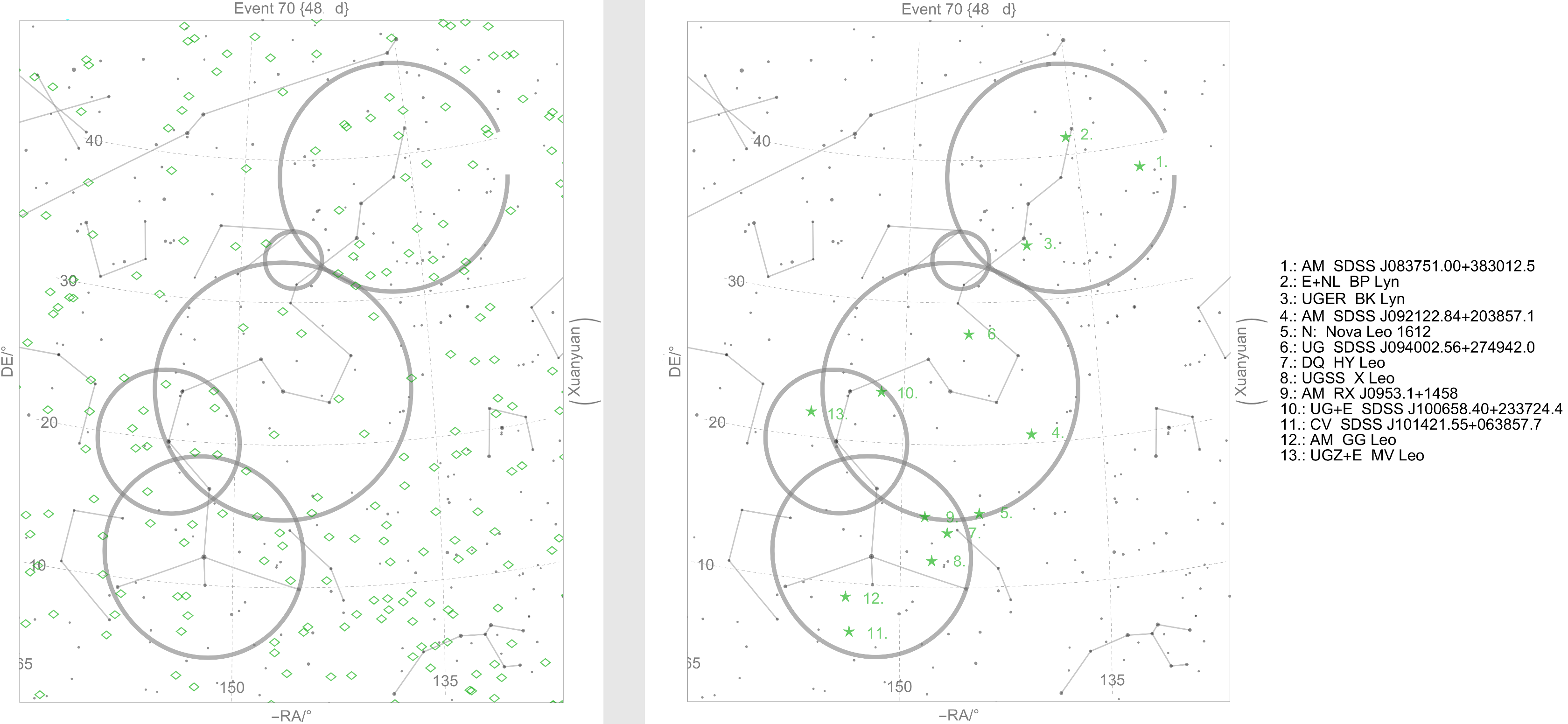}
    \caption{Event $+70$.}
    \label{fig:event70}
\end{figure*}

\begin{figure*}
	\includegraphics[width=\textwidth]{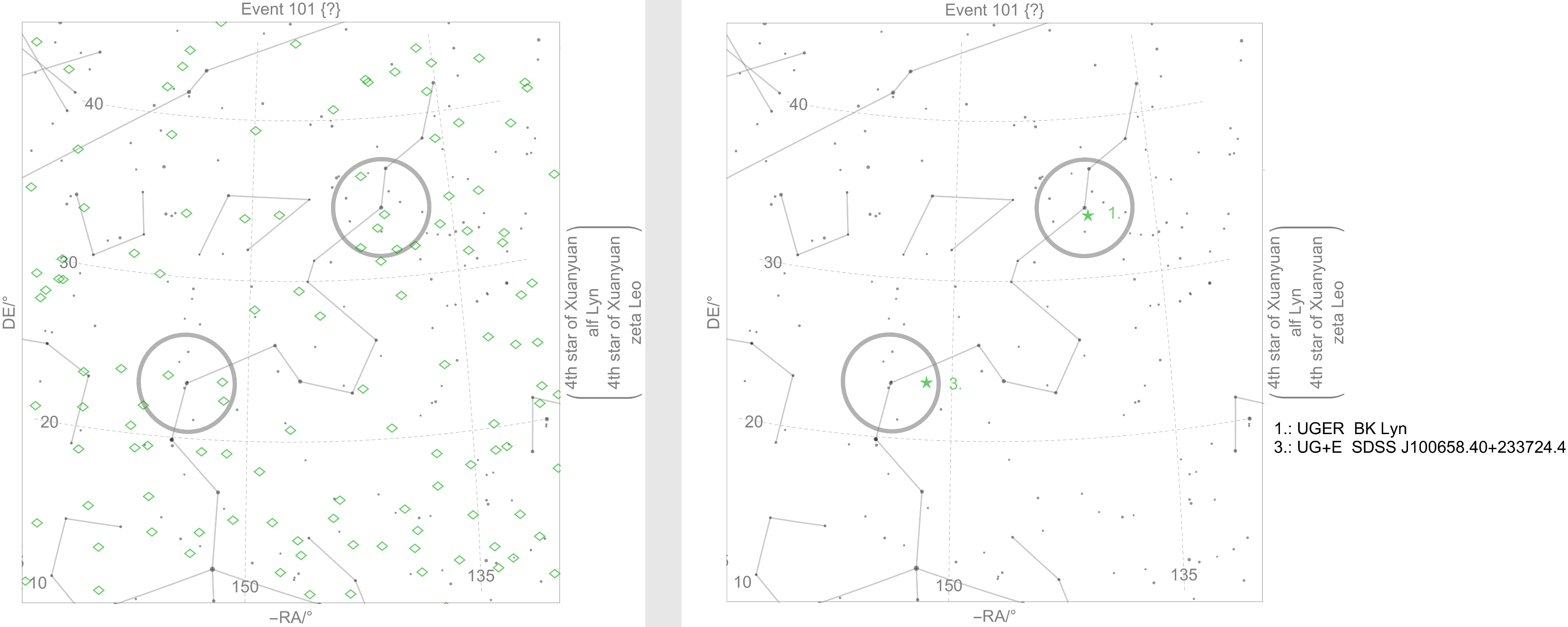}
    \caption{Event $+101$.}
    \label{fig:event101}
\end{figure*}

\begin{figure*}
	\includegraphics[width=\textwidth]{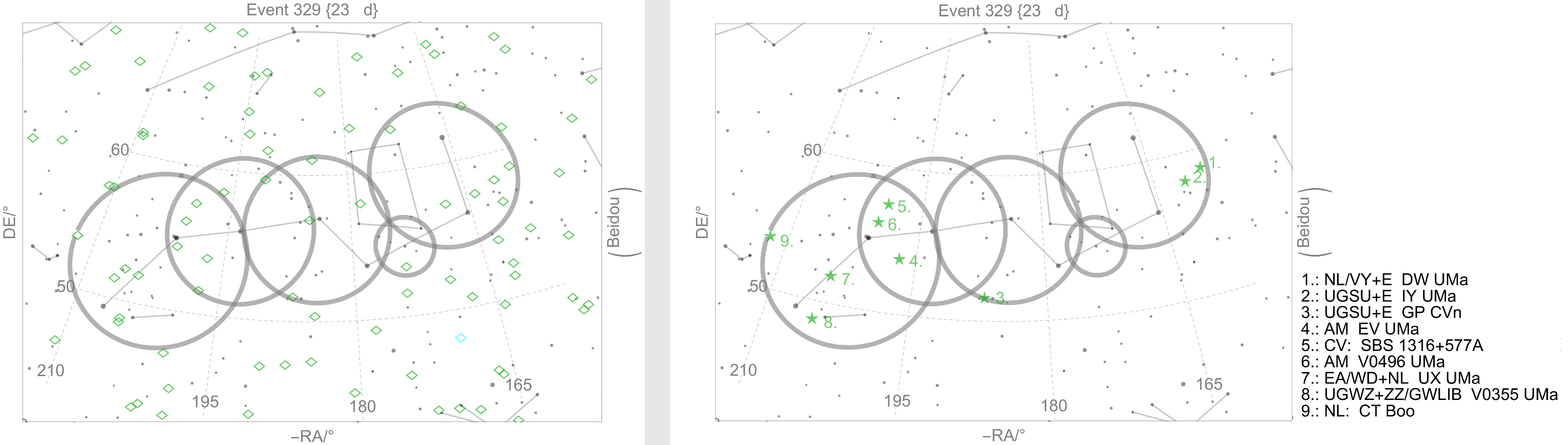}
    \caption{Event $+329$.}
    \label{fig:event329}
\end{figure*}

\begin{figure*}
	\includegraphics[width=\textwidth]{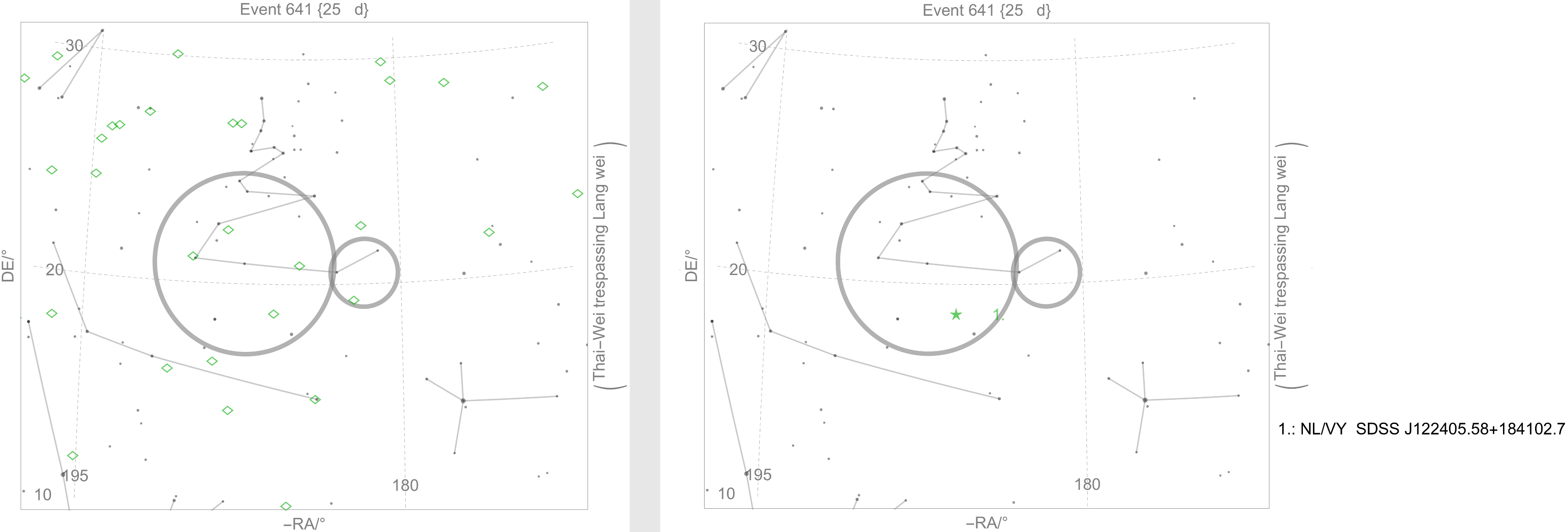}
    \caption{Event $+641$.}
    \label{fig:event641}
\end{figure*}

\begin{figure*}
	\includegraphics[width=\textwidth]{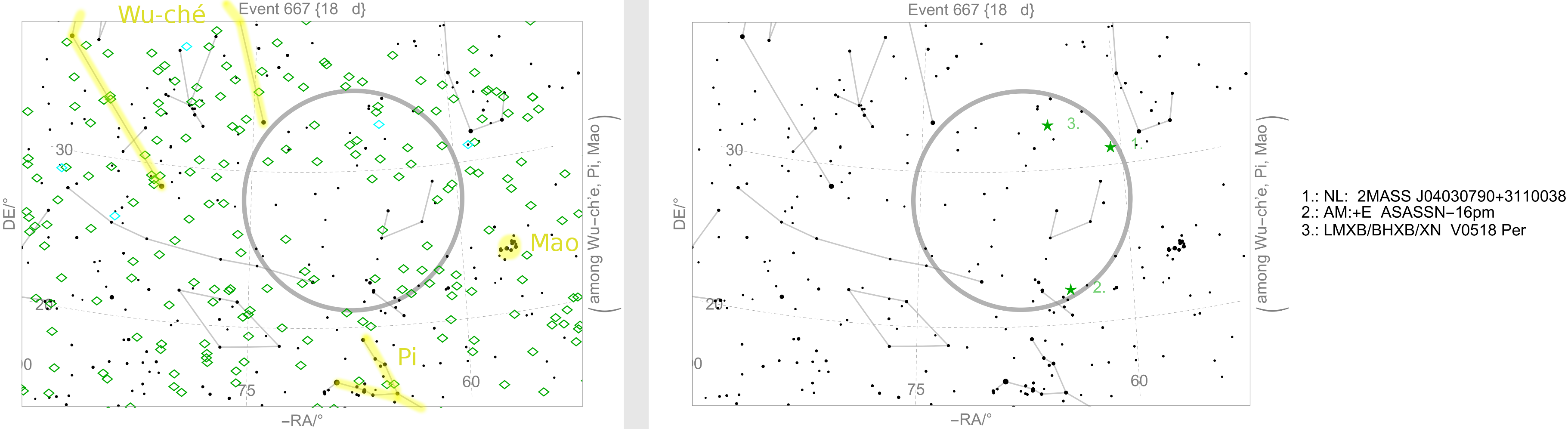}
    \caption{Event $+667$.}
    \label{fig:event667}
\end{figure*}

\begin{figure*}
	\includegraphics[width=\textwidth]{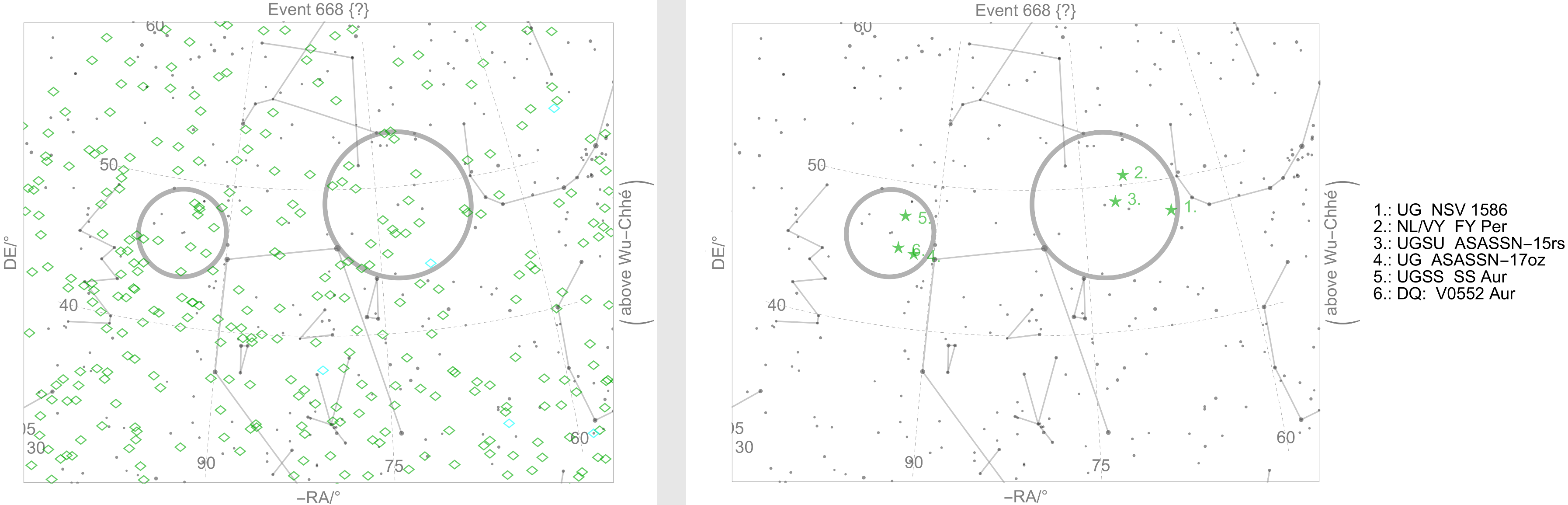}
    \caption{Event $+668$.}
    \label{fig:event668}
\end{figure*}

\begin{figure*}
	\includegraphics[width=\textwidth]{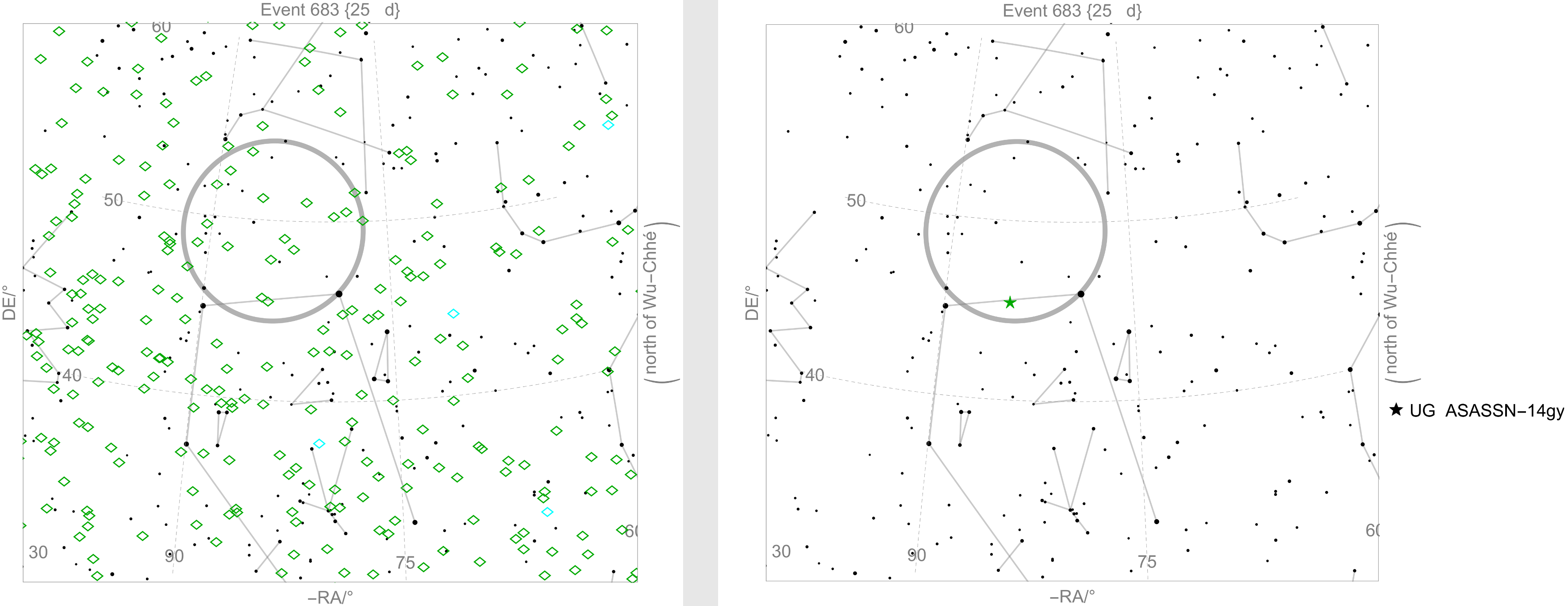}
    \caption{Event $+683$.}
    \label{fig:event683}
\end{figure*}

\begin{figure*}
	\includegraphics[width=\textwidth]{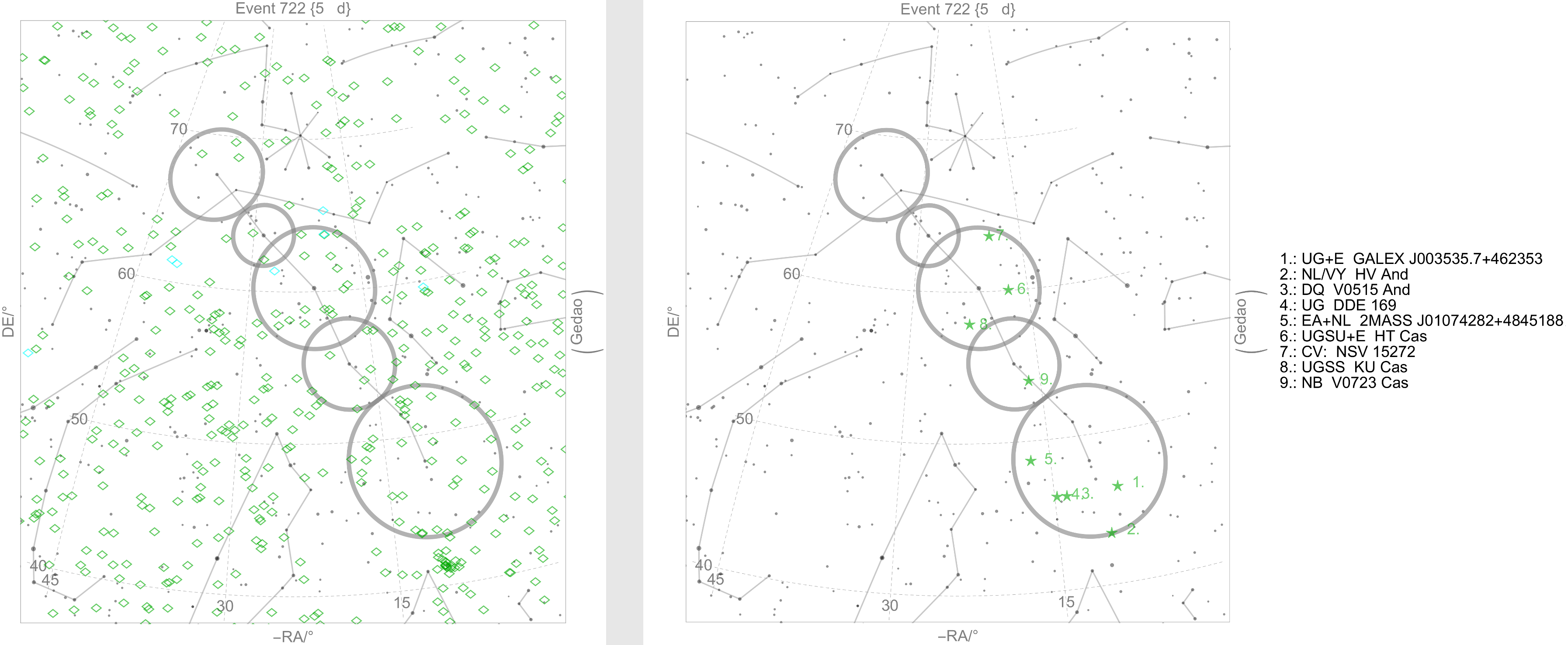}
    \caption{Event $+722$.}
    \label{fig:event722}
\end{figure*}

\begin{figure*}
	\includegraphics[width=\textwidth]{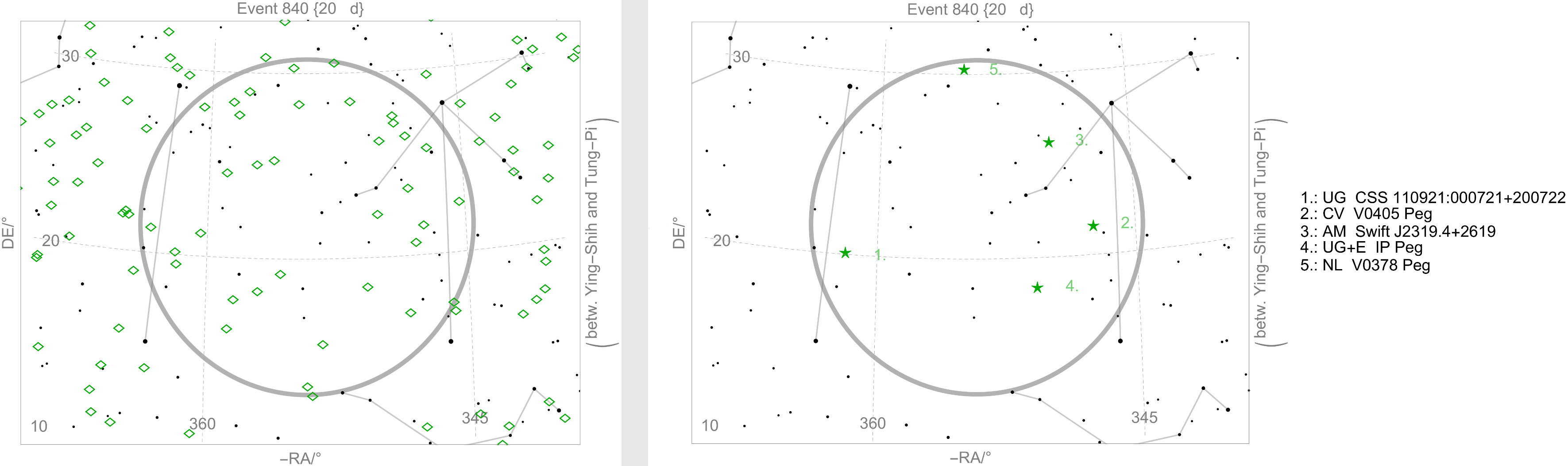}
    \caption{Event $+840$.}
    \label{fig:event840}
\end{figure*}

\begin{figure*}
	\includegraphics[width=\textwidth]{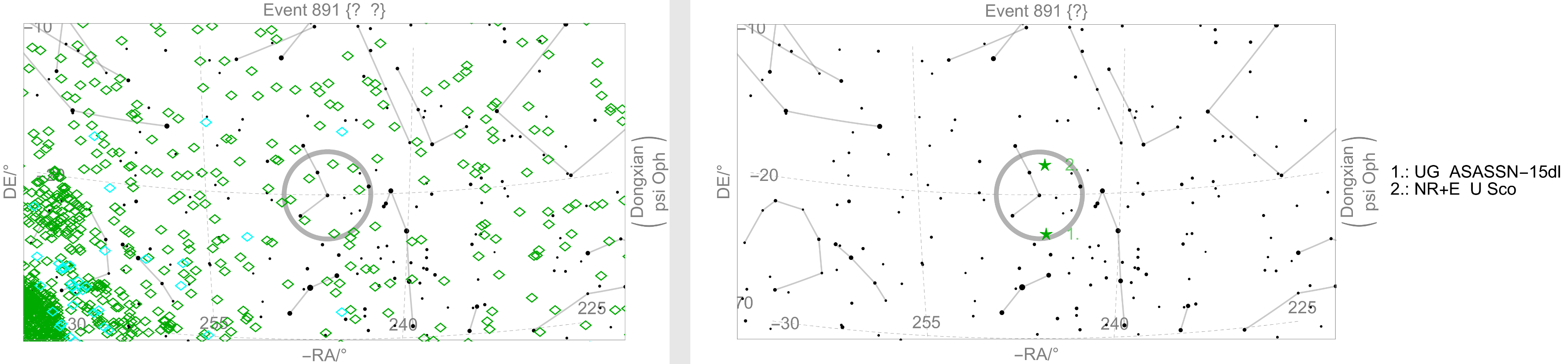}
    \caption{Event $+891$.}
    \label{fig:event891}
\end{figure*}

\begin{figure*}
	\includegraphics[width=\textwidth]{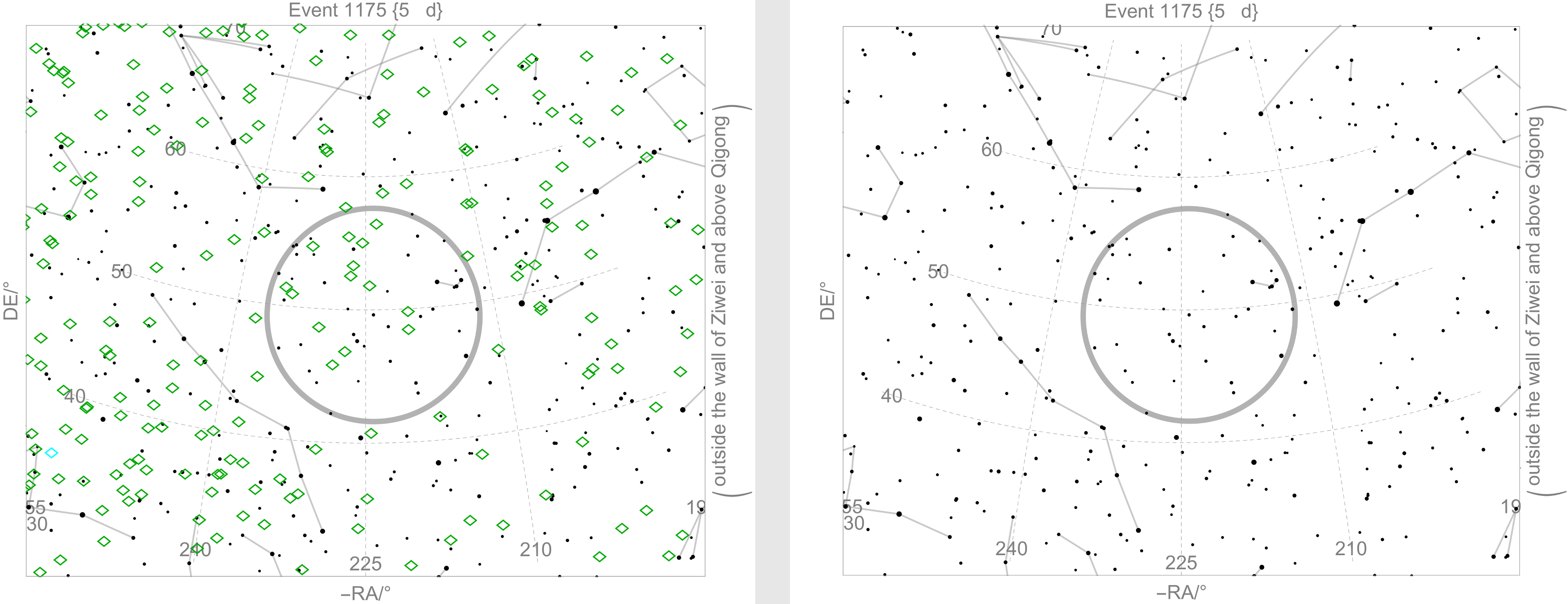}
    \caption{Event $1175$.}
    \label{fig:event1175}
\end{figure*}

\begin{figure*}
	\includegraphics[width=\textwidth]{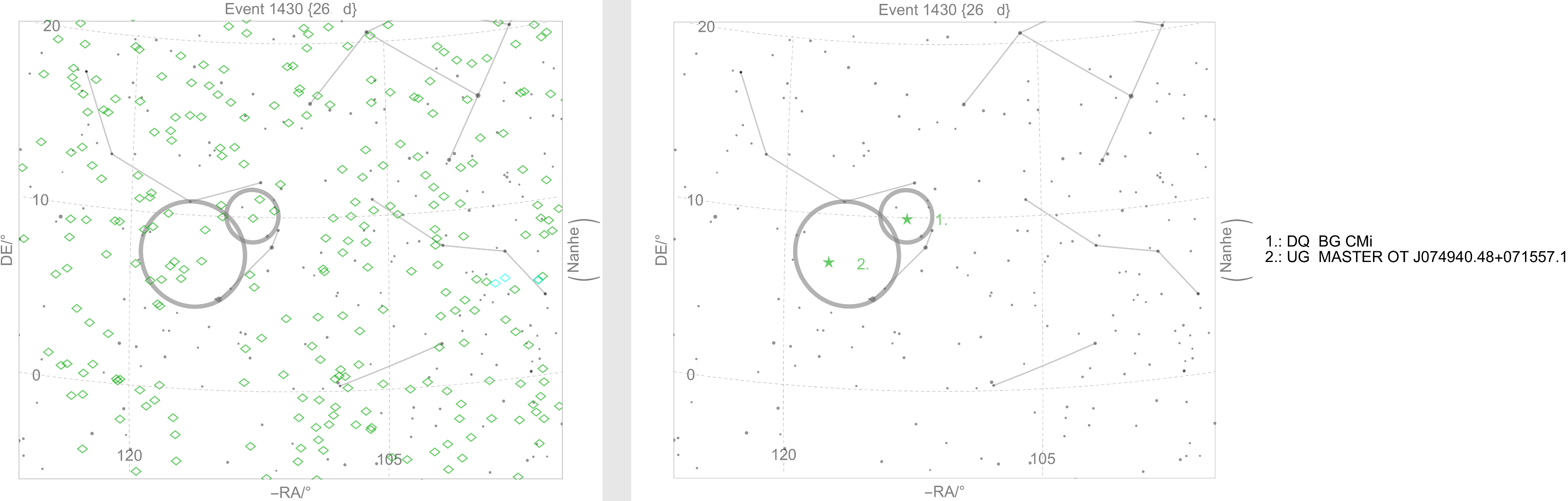}
    \caption{Event $1430$.}
    \label{fig:event1430}
\end{figure*}

\begin{figure*}
	\includegraphics[width=\textwidth]{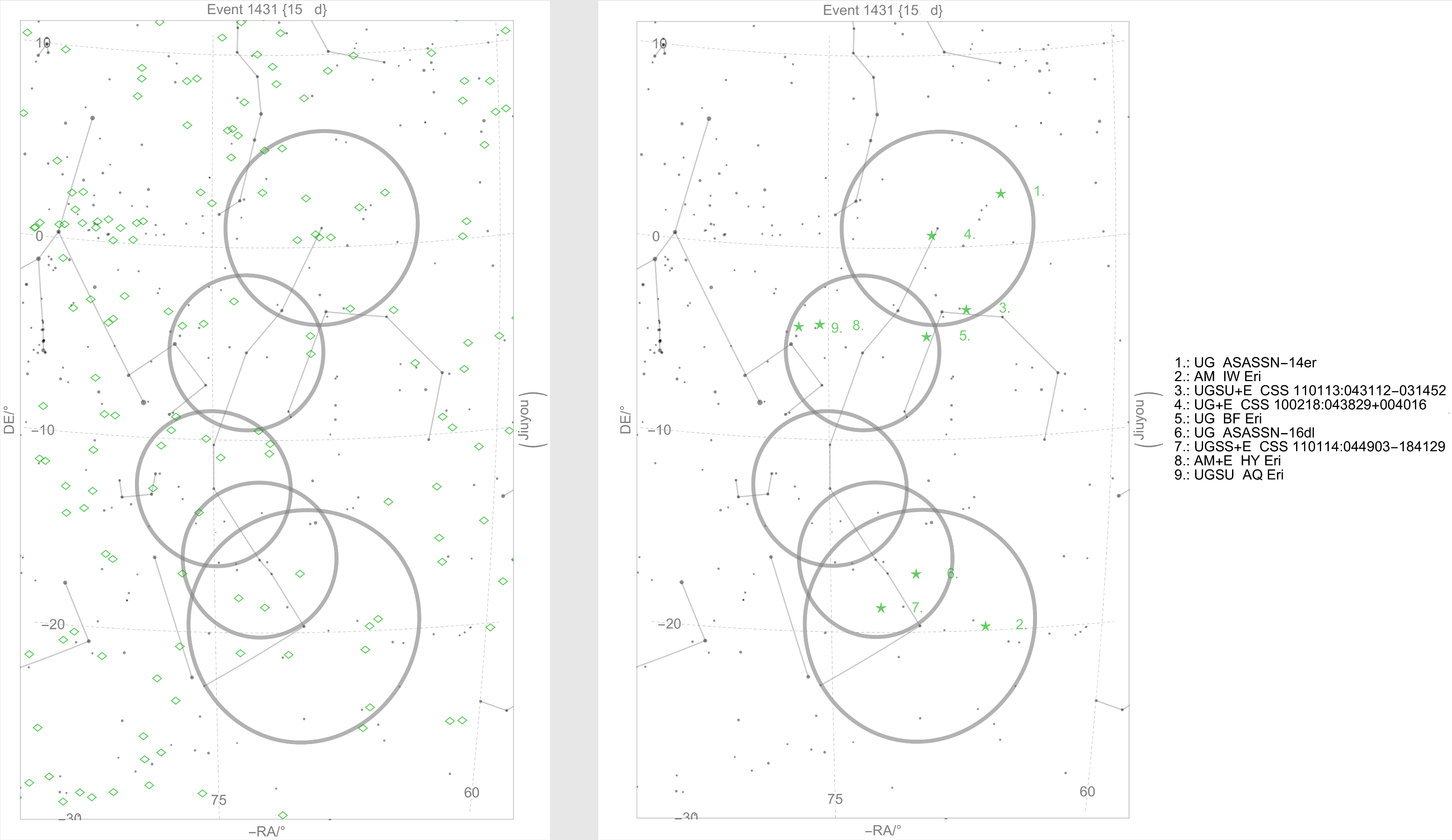}
    \caption{Event 1431.}
    \label{fig:event1431}
\end{figure*}

\begin{figure*}
	\includegraphics[width=\textwidth]{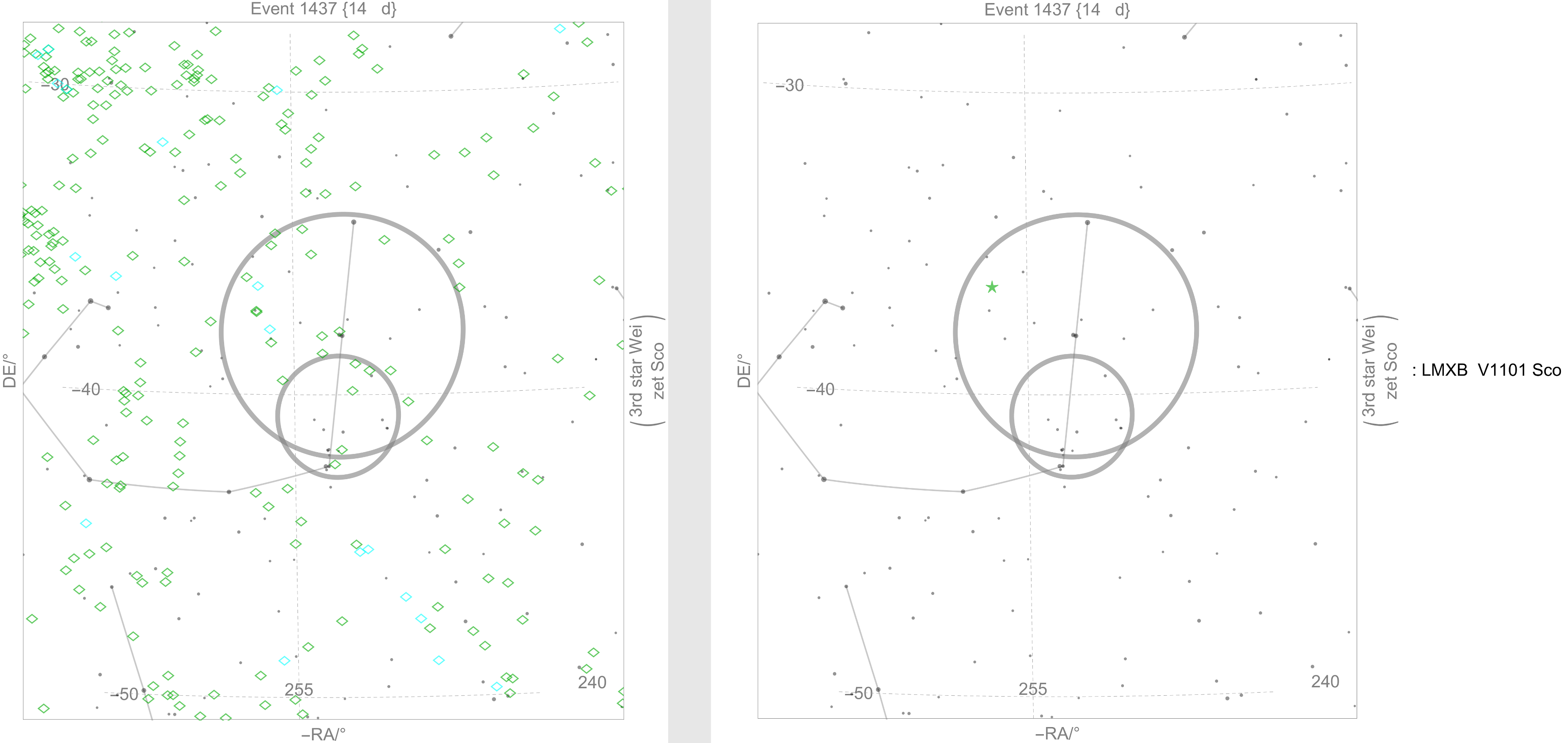}
    \caption{Event 1437. Obviously, the only candidate is not between the 2nd and 3rd star of the chain and must be dropped.}
    \label{fig:event1437}
\end{figure*}

\begin{figure*}
	\includegraphics[width=\textwidth]{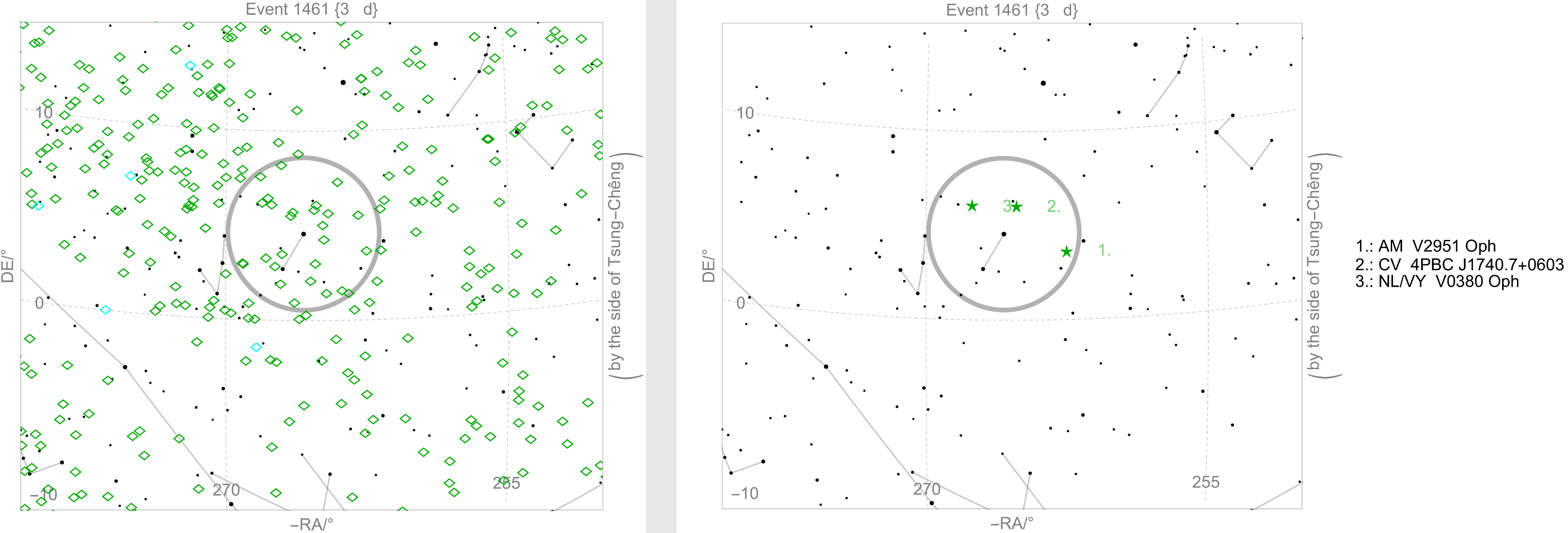}
    \caption{Event 1461.}
    \label{fig:event1461}
\end{figure*}

\begin{figure*}
	\includegraphics[width=\textwidth]{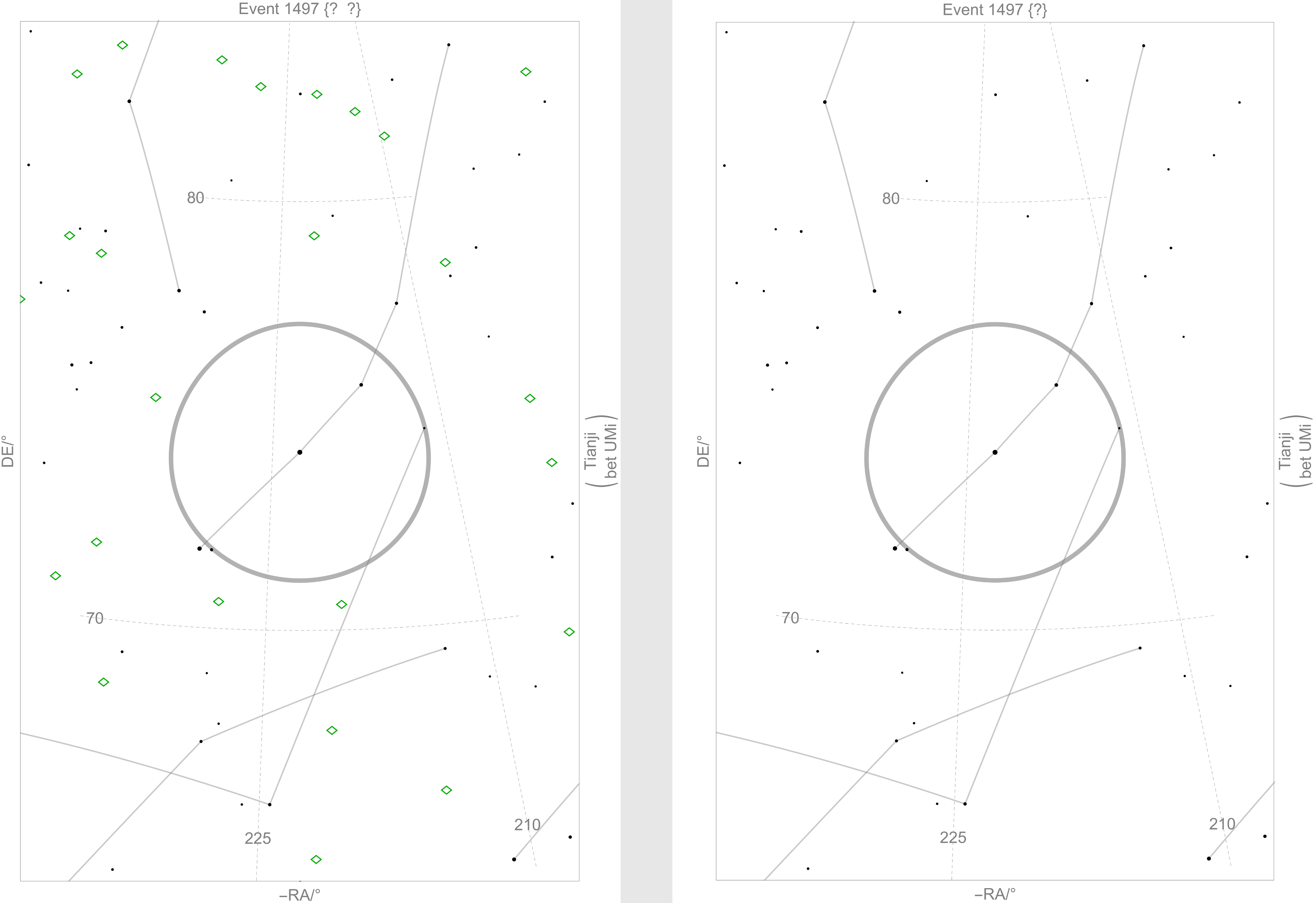}
    \caption{Event 1497.}
    \label{fig:event1497}
\end{figure*}

\begin{figure*}
	\includegraphics[width=\textwidth]{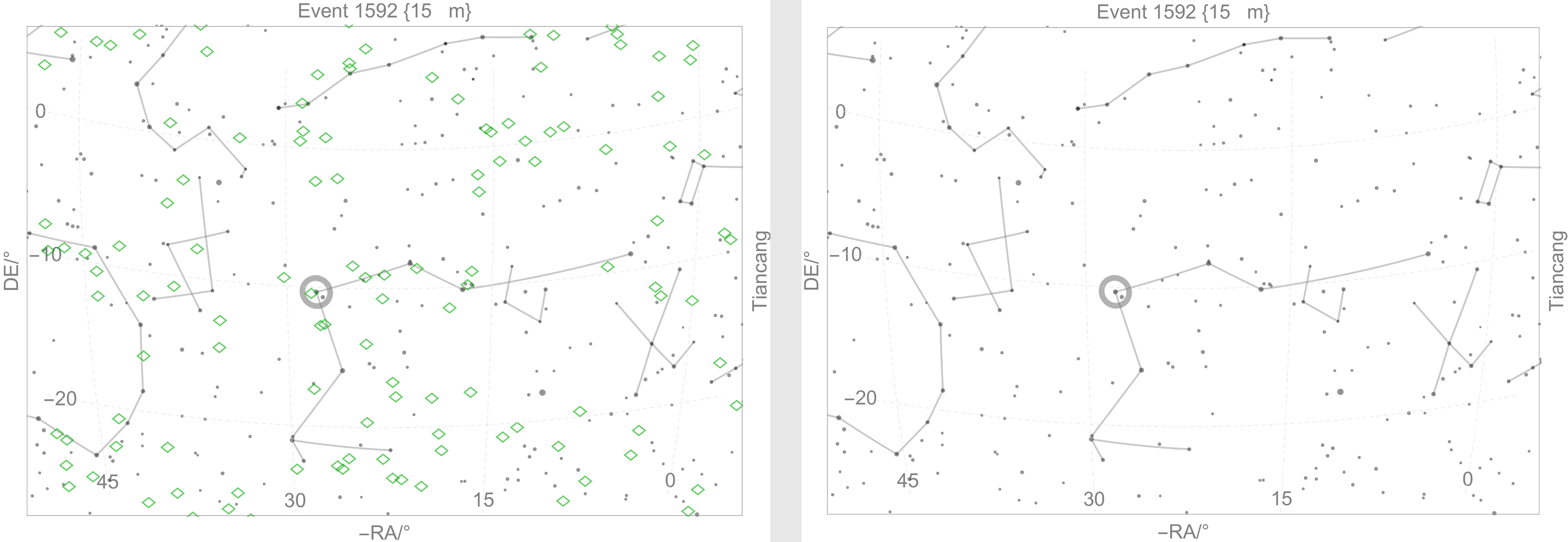}
    \caption{Event 1592 in Tiancang.}
    \label{fig:event1592T}
\end{figure*}

\begin{figure*}
	\includegraphics[width=\textwidth]{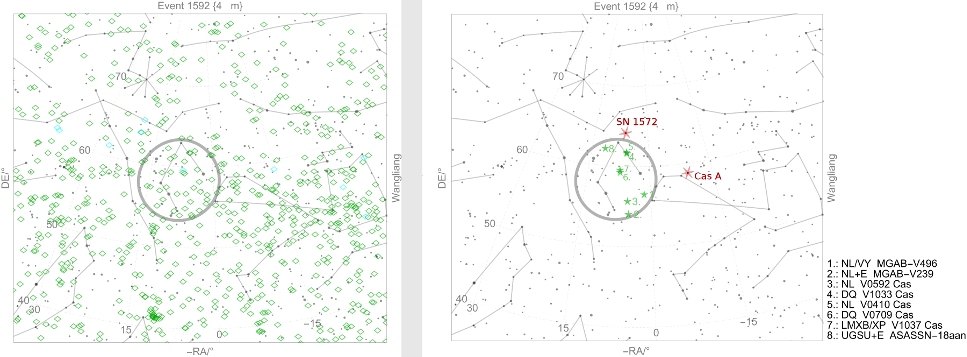}
    \caption{Event 1592 in Wangliang.}
    \label{fig:event1592W}
\end{figure*}

\begin{figure*}
	\includegraphics[width=\textwidth]{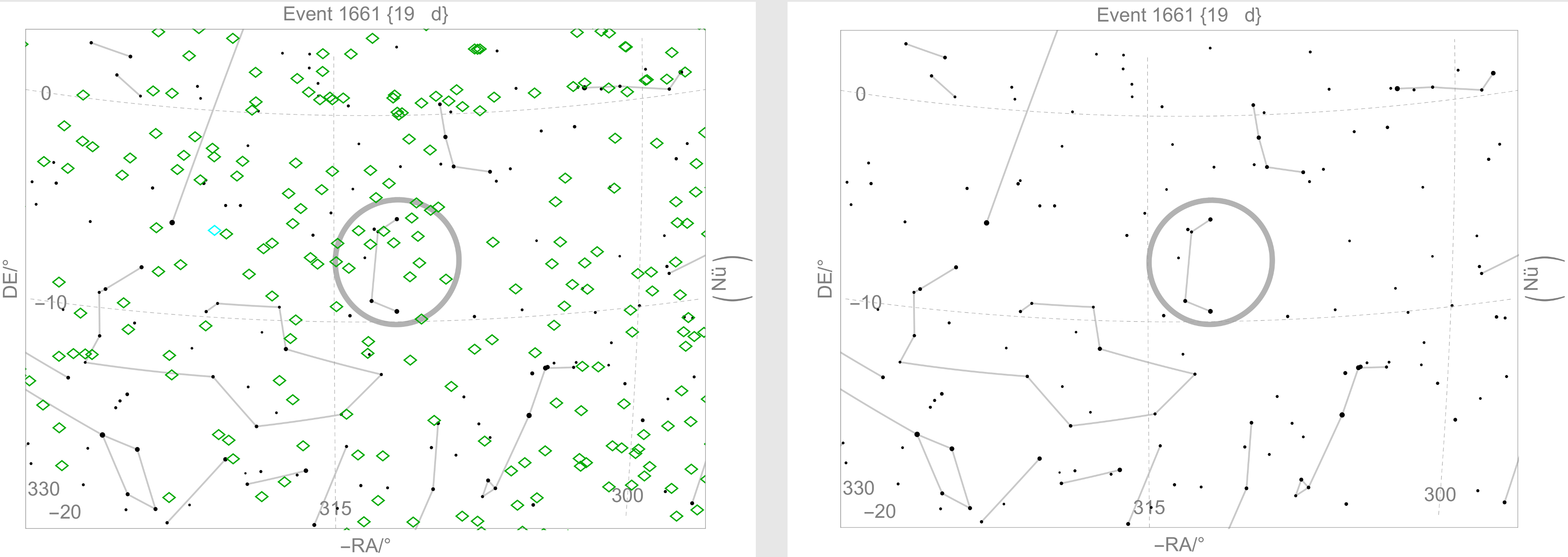}
    \caption{Event 1661.}
    \label{fig:event1661}
\end{figure*}

\begin{figure*}
	\includegraphics[width=\textwidth]{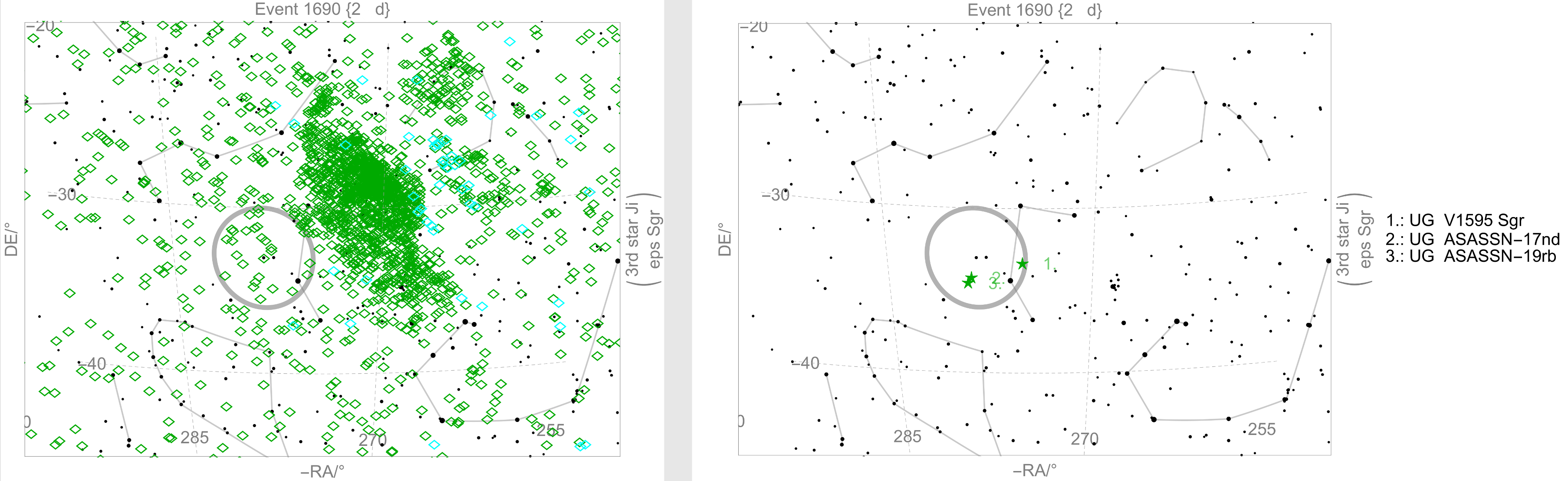}
    \caption{Event 1690.}
    \label{fig:event1690}
\end{figure*}

\bsp	
\label{lastpage}
\end{document}